\documentclass[a4paper,10pt]{article}
\usepackage[english]{babel}
\usepackage{jheppub}
\usepackage[T1]{fontenc} 
\usepackage{graphicx}
\usepackage{amsmath}
\usepackage{color}
\usepackage{xpatch}
\usepackage{caption}
\usepackage{subcaption}
\usepackage{slashed}
\usepackage{cleveref}
\usepackage{tikz}
\usetikzlibrary{decorations.pathreplacing}
\usepackage[normalem]{ulem}

\renewcommand{\i}{\ensuremath{\mathrm{i}}}
\renewcommand{\d}{\ensuremath{\mathrm{d}}}
\newcommand{\eps}{\ensuremath{\epsilon}}

\xpretocmd{\eqref}{eq.~}{}{}
\newcommand{\tb}{{\bar{t}}}
\newcommand{\as}{{\alpha_s}}
\newcommand{\epsv}{\varepsilon}
\def\vareps{\varepsilon}
\newcommand{\GammaBold}{\mathbf{\Gamma}}
\newcommand{\GammaPrime}{\Gamma^\prime}
\def\cT{\mathcal{T}}
\def\cF{\mathcal{F}}
\def\cO{\mathcal{O}}
\def\cR{\mathcal{R}}
\def\ZZ{\mathbf{Z}}
\def\dZ{\delta Z}
\def\dk#1{\frac{\mathrm{d}^d k_{#1}}{\mathrm{i} \pi^{d/2} \mathrm{e}^{-\eps \gamma_E}}}

\def\la{\langle}
\def\ra{\rangle}
\def\spA#1#2{\la#1#2\ra}
\def\spB#1#2{[#1#2]}

\def\spAA#1#2#3{\la#1|#2|#3\ra}
\def\spBB#1#2#3{[#1|#2|#3]}

\def\sqrtsymb{\sqrt{S}}
\def\mren{\mathrm{mren}}

\allowdisplaybreaks[4]

\title{Numerical evaluation of two-loop QCD helicity amplitudes for $gg\to t\tb g$ at leading colour}

\author[a]{Simon Badger,}
\author[b]{Matteo Becchetti,}
\author[a]{Colomba Brancaccio,}
\author[c,d]{Heribertus Bayu Hartanto,}
\author[e]{Simone Zoia}
\affiliation[a]{Dipartimento di Fisica and Arnold-Regge Center, Università di Torino, and INFN, Sezione di Torino,
Via P.\ Giuria 1, I-10125 Torino, Italy}
\affiliation[b]{Dipartimento di Fisica e Astronomia, Università di Bologna e INFN, Sezione di Bologna, via Irnerio 46,
I-40126 Bologna, Italy}
\affiliation[c]{Asia Pacific Center for Theoretical Physics, Pohang, 37673, Korea}
\affiliation[d]{Department of Physics, Pohang University of Science and Technology, Pohang, 37673, Korea}
\affiliation[e]{Physik-Institut, Universität Zürich, Winterthurerstrasse 190, 8057 Zürich, Switzerland}

\emailAdd{simondavid.badger@unito.it, matteo.becchetti@unibo.it, colomba.brancaccio@unito.it, bayu.hartanto@apctp.org, simone.zoia@physik.uzh.ch}
 
\preprint{ZU-TH-65/24}
 
\abstract{
We present the first benchmark evaluation of the two-loop finite remainders
for the production of a top-quark pair in association with a jet at hadron colliders in the
gluon channel. We work in the leading colour approximation, and perform the numerical
evaluation in the physical phase space. To achieve this result, we develop a new 
method for expressing the master integrals in terms of 
a (over-complete) basis of special functions that enables the infrared and ultraviolet 
poles to be cancelled analytically despite the presence of elliptic Feynman integrals. 
The special function basis makes it manifest that the elliptic functions appear solely
in the finite remainder, and can be evaluated numerically through generalised series 
expansions. The helicity amplitudes are constructed using four dimensional projectors combined with finite-field techniques to perform 
integration-by-parts reduction, mapping to special functions and Laurent expansion in the dimensional regularisation parameter.
}

\date{}
\begin{document}
\maketitle
\flushbottom

\section{Introduction}

The production of a top-quark pair in association with a jet is a high priority
process for precision studies at the Large Hadron Collider (LHC), particularly at its high-luminosity phase (HL-LHC). 
The sensitivity to fundamental parameters of the Standard Model and the increasing precision of experimental data demand that this process is computed to at least next-to-next-to-leading order (NNLO) accuracy in QCD, for which theoretical challenges must be overcome.

A significant fraction (around 50\%) of all $t\bar{t}$ signals at the LHC are
associated with an additional QCD jet and have been extensively studied by the
ATLAS and CMS
experiments~\cite{CMS:2016oae,ATLAS:2018acq,CMS:2020grm,CMS:2024ybg}.
Normalised distributions for this event are highly sensitive to the top quark
mass and extensive studies have been made to develop a competitive parameter-extraction procedure~\cite{Alioli:2013mxa,Bevilacqua:2017ipv,Alioli:2022ttk,Alioli:2022lqo}.
Next-to-leading order (NLO) theory predictions have been available for nearly
20 years~\cite{Dittmaier:2007wz} and are now simple to compute using modern
automated tools such as \textsc{OpenLoops2}~\cite{Buccioni:2019sur}, \textsc{Helac-NLO}\cite{Bevilacqua:2011xh} or
\textsc{MadGraph5\_aMC@NLO}~\cite{Alwall:2014hca}. The theory-level input to such studies is
currently at the level of NLO in QCD matched to a parton shower (NLO+PS), which has
been implemented within the \textsc{Powheg-Box} framework~\cite{Alioli:2011as} with
complete off-shell effects also available at fixed
order~\cite{Bevilacqua:2015qha,Bevilacqua:2016jfk}. Furthermore, mixed QCD and electroweak (EW) corrections have also been calculated~\cite{G_tschow_2018}. Extending theoretical predictions to NNLO QCD
accuracy presents a major challenge since one must overcome bottlenecks
associated with the large number of variables of a three-particle final state
and the analytic complexity of the Feynman integrals with internal masses. Such
difficulties mean that the two-loop amplitudes entering the double virtual
corrections to the cross section are still currently unknown. The use of finite
field reconstruction techniques~\cite{vonManteuffel:2014ixa,Peraro:2016wsq,Klappert:2019emp,Peraro:2019svx,Smirnov:2019qkx,Klappert:2020aqs,Klappert:2020nbg}
has had a dramatic effect on our ability to evaluate the double virtual
corrections to complicated $2\to 3$ scattering processes with massless internal
particles~\cite{Badger:2019djh,Badger:2021nhg,Badger:2021ega,Abreu:2021asb,Badger:2022ncb,Agarwal:2021vdh,Badger:2021imn,Abreu:2023bdp,Badger:2023mgf,Agarwal:2023suw,DeLaurentis:2023nss,DeLaurentis:2023izi}, with up to one additional external scale~\cite{Badger:2021nhg,Badger:2021ega,Abreu:2021asb,Badger:2022ncb,Badger:2024sqv,Badger:2024awe} and including internal massive propagators \cite{Agarwal:2024jyq}. Progress has also required the development of highly optimised
integration-by-parts reduction techniques~\cite{Gluza:2010ws,Ita:2015tya,Larsen:2015ped,Wu:2023upw} and improved analytic understanding of
multi-scale Feynman integrals through their differential equations~\cite{Gehrmann:2015bfy,Papadopoulos:2015jft,Abreu:2018rcw,Chicherin:2018mue,Chicherin:2018old,Abreu:2018aqd,Abreu:2020jxa,Canko:2020ylt,Abreu:2021smk,Kardos:2022tpo,Abreu:2023rco} leading to numerically efficient bases of special functions referred to as \textit{pentagon functions}~\cite{Gehrmann:2018yef,Chicherin:2020oor,Chicherin:2021dyp,Abreu:2023rco,Gehrmann:2024tds,FebresCordero:2023pww}. The new results have been combined with unresolved radiation contributions to provide differential distributions at NNLO QCD for a variety of important experimental measurements~\cite{Chawdhry:2019bji,Kallweit:2020gcp,Chawdhry:2021hkp,Czakon:2021mjy,Badger:2021ohm,Chen:2022ktf,Alvarez:2023fhi,Badger:2023mgf,Hartanto:2022qhh,Hartanto:2022ypo,Buonocore:2022pqq,Catani:2022mfv,Buonocore:2023ljm,Mazzitelli:2024ura,Devoto:2024nhl,Biello:2024pgo}.

In this article we continue the efforts to compute the double virtual
corrections at NNLO. To this point, the corrections at one-loop order expanded to
$\mathcal{O}(\eps^2)$ have been considered~\cite{Badger:2022mrb} and the differential equations (DEs) for the
two-loop master integrals in the leading colour limit have been recently
completed~\cite{Badger:2022hno,Badger:2024fgb}. A major difference compared to other amplitudes with $2\to 3$ kinematics that have been previously studied is the presence of \emph{elliptic functions} in the master integrals~\cite{Bourjaily:2022bwx}. This feature makes the computation more challenging than in the previous cases.
In scenarios where all propagators are massless, the DEs were cast in a canonical form~\cite{Henn:2013pwa}, where the dimensional regulator factorises and the connection matrix contains solely logarithmic one-forms (``$\d \log$'s'').
This form of the DEs could then be solved in terms of a basis of polylogarithmic special functions known as \emph{pentagon functions}~\cite{Gehrmann:2018yef,Chicherin:2020oor,Badger:2021nhg,Chicherin:2021dyp,Abreu:2023rco}.
The latter are algebraically independent---in this sense they are a `basis'---and thus give a unique representation of the amplitudes where cancellations and simplifications are manifest.
Furthermore, they are both fast and stable for numerical evaluation.
Both of these advantages are spoiled by the presence of elliptic functions.
First, the fact that we do not have a canonical form of the DEs means that we cannot write the solution in terms of a basis of special functions.
This is a problem because the algebraic complexity of the amplitude at the
level of the master integral is significantly greater than after the expansion in the dimensional regulator and resolution of the functional relations. Second, the numerical evaluation of the master integrals in this case is substantially less efficient than in the fully canonical scenario~\cite{Badger:2024fgb}.

To address these challenges, we propose a strategy aimed at minimising the impact of the non-polylogarithmic functions.
We solve the non-canonical DEs for the master integrals in terms of a set of special functions that is potentially over-complete, 
but that nonetheless allows us to subtract analytically the ultraviolet (UV) and infrared (IR) poles of the two-loop amplitudes, and unlocks important simplifications in the latter. 
The large majority of the special functions are polylogarithmic and are thus suitable to be evaluated with the method applied to the pentagon functions. 
For the few non-polylogarithmic functions, we set up a minimal system of DEs which we solve by means of generalised power series~\cite{Moriello:2019yhu,Hidding:2020ytt}.

We employ this special-function representation of the master integrals in the computation of the two-loop leading-colour amplitude for  the production of two on-shell top quarks and a gluon in the gluon-fusion channel ($gg \to t\bar{t}g$) without closed fermion loops.
From the amplitude point of view, this is the most complicated partonic channel contributing to $t\bar{t}$ production in association with a jet at hadron colliders.
Obtaining analytic expressions for this amplitude is a huge challenge on its own due to the high algebraic complexity of the rational functions accompanying the master integrals or special functions.
We set up a routine based on finite-field arithmetic to evaluate numerically the two-loop amplitude, while concurrently also assess different strategies to derive its analytic form.
We employ the four-dimensional projection method~\cite{Peraro:2019cjj,Peraro:2020sfm,Buccioni:2023okz} to derive the helicity amplitudes and present benchmark numerical evaluations in physical kinematics.

Our paper is organised as follows: We begin in \cref{sec:struc} by
outlining the structure of the leading colour amplitude, its pole structure and
definition of the finite remainder. In \cref{sec:proj} we describe the
application of the four dimensional projector method to compute helicity
amplitudes suitable for the description of top-quark decays in the narrow width
approximation. In \cref{sec:special_functions} we discuss the basis of functions
used to describe the finite remainder before presenting benchmark evaluations
and cross-checks in \cref{sec:res}. We conclude with a summary and
remarks regarding the prospects for the derivation of amplitudes suitable for
phenomenological applications.

\section{Colour and pole structure of the leading colour amplitude \label{sec:struc}}

We consider the partonic process
\begin{equation} \label{eq:Process}
	g(-p_4) + g(-p_5) \to \bar{t}(p_1) + t(p_2) + g(p_3) \,.
\end{equation}
We work in the ’t Hooft-Veltman (HV) scheme with $d = 4 - 2 \eps$ space-time dimensions and four-dimensional external momenta $p_i$.
The latter are all outgoing and satisfy the following momentum-conservation and on-shell conditions,
\begin{equation} 
  p_1+p_2+p_3+p_4+p_5 = 0 \,, \qquad \quad
	p_1^2 = p_2^2 = m_t^2, \qquad \quad p_3^2 = p_4^2 = p_5^2 = 0 \,, 
\end{equation}
where \(m_t\) is the top-quark mass.
The kinematics are described by six scalar invariants, which we choose as
\begin{align}
\label{eq:dijmtset}
\vec{d} = \bigl( d_{12}, d_{23}, d_{34}, d_{45}, d_{15}, m_t^2 \bigr) \,,
\end{align}
with $d_{ij} = p_i \cdot p_j$, and one pseudo-scalar invariant,
\begin{equation}
	\mathrm{tr}_5 = 4 \, \i \, \varepsilon_{\mu\nu\rho\sigma} \, p_1^{\mu} p_2^{\nu} p_3^{\rho} p_4^{\sigma} \,,
\end{equation}
where $\varepsilon_{\mu\nu\rho\sigma}$ is the anti-symmetric Levi-Civita pseudo-tensor.

We compute the two-loop amplitude in the limit of a large number of colours $N_c$,
neglecting the contributions from closed fermion loops.
Only planar diagrams contribute at the leading order in this limit, resulting in a simpler set of Feynman integrals. 
The leading colour $L$-loop bare amplitude $\mathcal{A}^{(L)}_{\rm LC}$ has the following colour decomposition in terms of partial amplitudes $A^{(L)}$, 
\begin{align} \label{eq:ColDec}
\begin{aligned} 
	\mathcal{A}^{(L)}_{\rm LC}\big(1_{\bar{t}},\, & 2_t, 3_g, 4_g, 5_g \big) = \sqrt{2}\;\overline{g}_s^3 \,n^L
	\sum_{\sigma \in Z_3} \left(t^{a_{\sigma(3)}} t^{a_{\sigma(4)}} t^{a_{\sigma(5)}}
	\right)^{\bar{i}_1}_{\ i_2} 
	A^{(L)} \bigl(1_{\bar{t}},\, 2_t, \sigma(3)_g, \sigma(4)_g, \sigma(5)_g \bigr)
	 \,,
\end{aligned}
\end{align} 
where $\overline{g}_s$ is the bare strong coupling constant, $n =N_{c} (4 \pi)^{\eps} \mathrm{e}^{- \eps \gamma_E}  \overline{\alpha}_s/(4 \pi)$, $\overline{\alpha}_s = \overline{g}_s^2/(4 \pi)$, $Z_3$ is the set of cyclic permutations of $(3,4,5)$, $ t^{a_i} $ are the fundamental generators of the \(SU(N_c)\) group normalised according to $\mathrm{tr}(t^a t^b) = \delta^{ab}/2$, \( a_i = 1, \dots, 8 \) index the adjoint representation, and \( i_2 \,(\bar{i}_1) = 1, 2, 3 \) index the fundamental (anti-fundamental) representation.

\begin{figure}[t]
    \centering
    \begin{subfigure}[b]{0.3\textwidth}
        \centering
        \includegraphics[height=4cm]{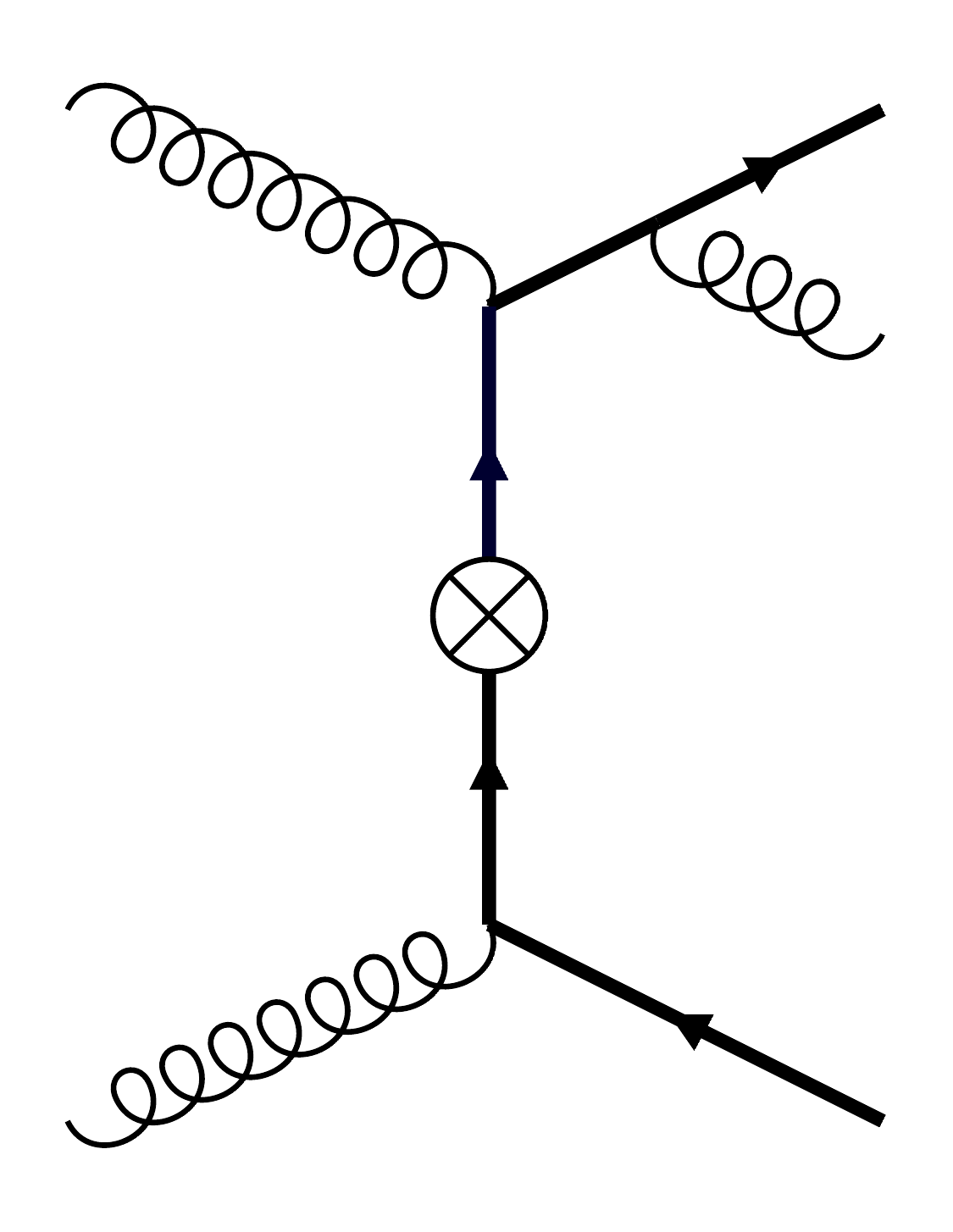}
        \caption{$A^{(0)}_{\mathrm{mct}}$}
        \label{fig:A0mct}
    \end{subfigure}
    \hfill
    \begin{subfigure}[b]{0.3\textwidth}
        \centering
        \includegraphics[height=4cm]{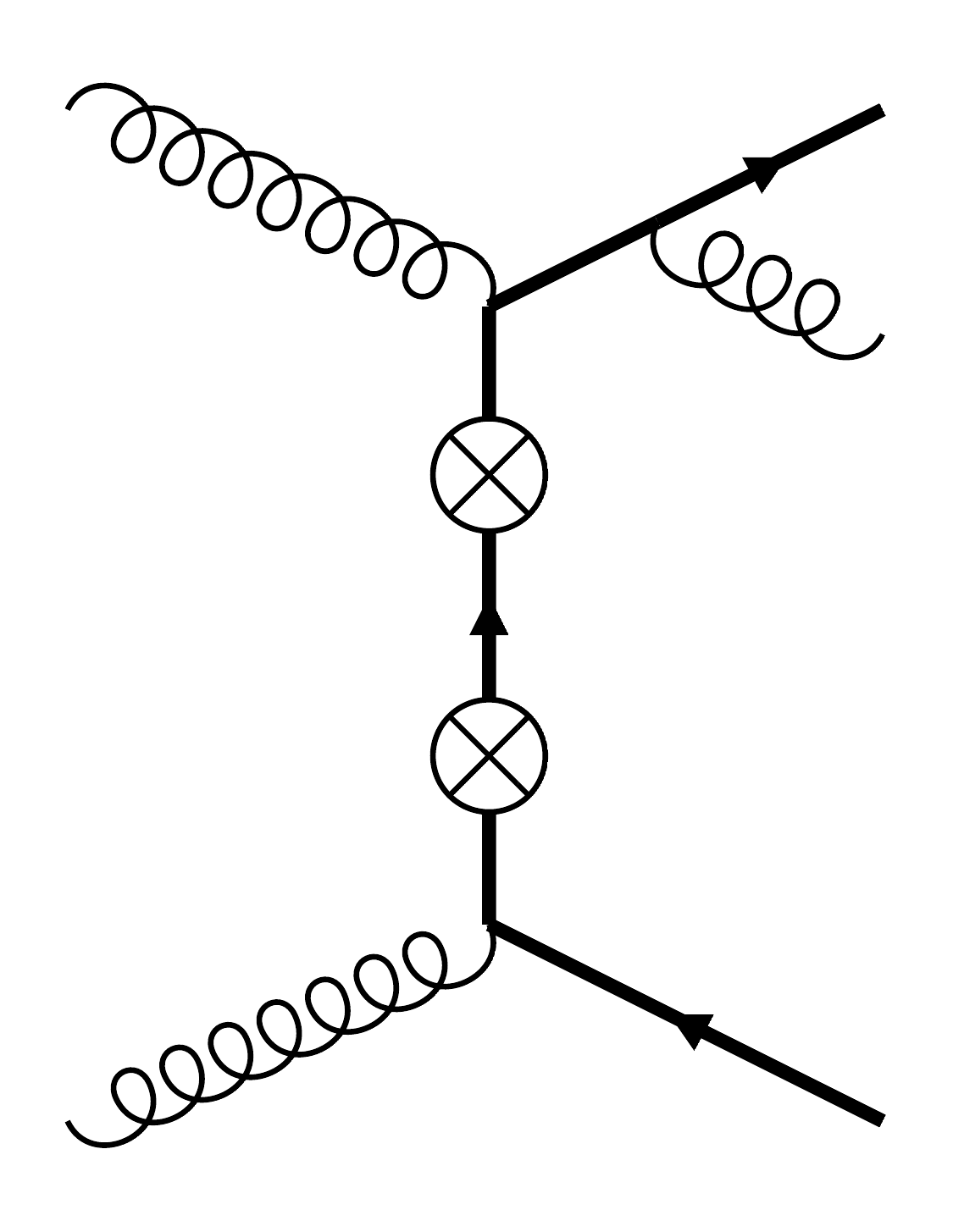}
        \caption{$A^{(0)}_{2\mathrm{mct}}$}
        \label{fig:A02mct}
    \end{subfigure}
    \hfill
    \begin{subfigure}[b]{0.3\textwidth}
        \centering
        \includegraphics[height=4cm]{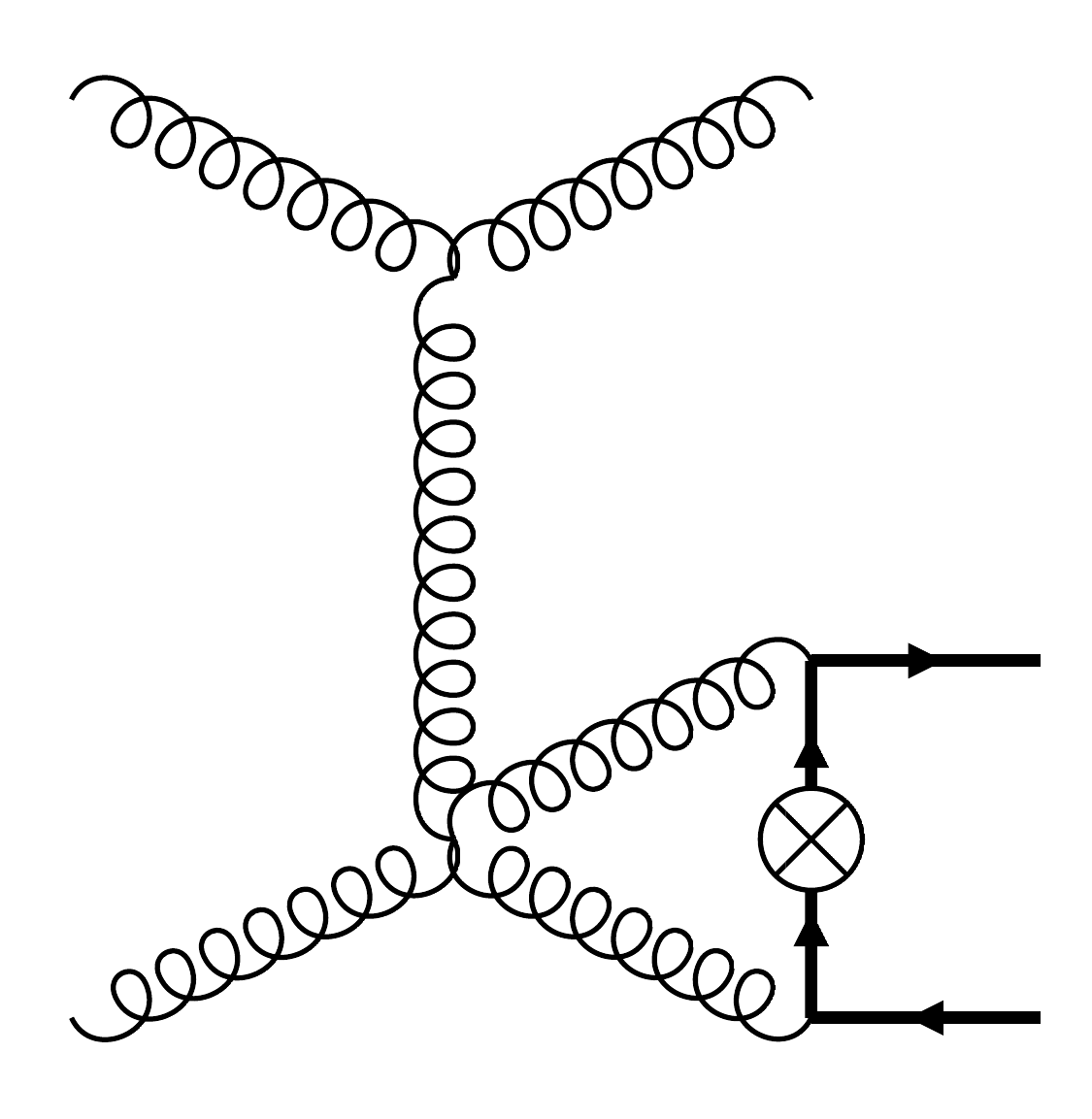}
        \caption{$A^{(1)}_{\mathrm{mct}}$}
        \label{fig:A1mct}
    \end{subfigure}

    \caption{Examples of Feynman diagrams with one and two mass-counterterm insertions, together with the terms of \cref{eq:Amren} they contribute to. Thick lines denote massive particles, and crossed circles indicate the counterterm insertions.}
    \label{fig:mct}
\end{figure}

We define gauge invariant one- and two-loop partial amplitudes by including the additive renormalisation of the top-quark mass, as
\begin{align} \label{eq:Amren}
\begin{aligned}
	A^{(1)}_{\mathrm{mren}} & = A^{(1)} - \dZ_m^{(1)} A^{(0)}_{\mathrm{mct}} \,, \\
	A^{(2)}_{\mathrm{mren}} & = A^{(2)} - \dZ_m^{(2)} A^{(0)}_{\mathrm{mct}} 
	+ (\dZ_m^{(1)})^2 A^{(0)}_{2\mathrm{mct}} - \dZ_m^{(1)} A^{(1)}_{\mathrm{mct}} \,.
\end{aligned}
\end{align}
Here, $A^{(0)}_{\mathrm{mct}}$ is the tree-level amplitude with the insertion of a single mass counterterm in the internal massive propagator, $A^{(0)}_{2\mathrm{mct}}$ is the tree-level amplitude with two mass-counterterm insertions, and $A^{(1)}_{\mathrm{mct}}$ is the one-loop amplitude with one mass-counterterm insertion. Examples of diagrams contributing to $A^{(0)}_{\mathrm{mct}}$, $A^{(0)}_{2\mathrm{mct}}$ and $A^{(1)}_{\mathrm{mct}}$ are shown in \cref{fig:mct}. Moreover, in \cref{eq:Amren}, $\dZ_m^{(L)}$ denotes the $L$-loop top-quark mass-renormalisation factors, for which a representation in terms of Feynman integrals was provided in Ref.~\cite{Badger:2021owl}. 
This form of the mass counterterms allows us to check the gauge invariance of the amplitudes already at the level of the master integrals. For completeness, we provide the expression of all renormalisation factors in \cref{app:renormalisation}.

In order to subtract the remaining UV singularities, we renormalise also the top-quark wavefunction and the strong coupling constant $\as$.
The counterterm associated with the gluon wavefunction contributes only to the amplitude terms arising from closed loops of heavy quarks~\cite{Mitov:2006xs, Czakon:2007wk}, which are excluded from our leading colour computation.
The renormalised partial amplitudes read
\begin{align} \label{eq:UVRen}
\begin{aligned}
	A^{(1)}_{\mathrm{ren}} = \ & A^{(1)}_{\mathrm{mren}}
	 + \left( \frac{3}{2} \dZ_{\as}^{(1)} + \dZ_t^{(1)} \right) A^{(0)} \,, \\
	A^{(2)}_{\mathrm{ren}} = \ & A^{(2)}_{\mathrm{mren}} + \left( 
	\frac{5}{2} \dZ_{\as}^{(1)} + \dZ_t^{(1)} \right) A^{(1)}_{\mathrm{mren}} \\
	& + \left( \frac{3}{2} \dZ_{\as}^{(2)} +\frac{5}{2} \dZ_{\as}^{(1)}
	\dZ_{t}^{(1)} + \dZ_t^{(2)} + \frac{3}{8} (\dZ_{\as}^{(1)})^2
	\right) A^{(0)} \,,
\end{aligned}
\end{align}
where the superscripts refer to the order in the renormalised strong coupling constant $\as$. 
Moreover, $A^{(1)}_{\mathrm{mren}}$ is the mass-renormalised one-loop amplitude defined in \eqref{eq:Amren} and $A^{(0)}$ is the tree-level amplitude. 
The $L$-loop renormalisation constants $\dZ^{(L)}_{t}$~\cite{Melnikov:2000zc} and $\dZ^{(L)}_{\as}$ are given in \cref{app:renormalisation}. 

The renormalised partial amplitudes still contain divergences of IR nature. 
The pole structure of these divergences can be predicted by universal formulae describing the IR behaviour of 
QCD amplitudes \cite{CATANI1998161, Becher:2009cu, Becher_2009, Gardi:2009qi, Gardi:2009zv, Catani:2000ef, Ferroglia:2009ep, Ferroglia:2009ii}. 
We define the partial finite remainders~as
\begin{align} \label{eq:IRRen}
\begin{aligned}
	R^{(0)} & = A^{(0)} \,, \\ 
	R^{(1)} & = A^{(1)}_{\mathrm{ren}} - \ZZ^{(1)} A^{(0)} \,, \\ 
	R^{(2)} & = A^{(2)}_{\mathrm{ren}} - \left( \ZZ^{(2)} - (\ZZ^{(1)})^2 \right) A^{(0)} 
	- \ZZ^{(1)} A^{(1)}_{\mathrm{ren}} \,.
\end{aligned}
\end{align}
The relation between the partial finite remainders $R^{(L)}$ and the colour-dressed finite remainder $\cR^{(L)}$ at leading colour takes the same form as for the 
bare amplitude, shown in \cref{eq:ColDec}, upon replacing $\mathcal{A} \rightarrow \cR$ and $A\rightarrow R$.
The expression of the IR pole operators $\ZZ^{(L)}$ are given in \cref{app:renormalisation}.

\section{Helicity amplitudes for $gg\to t\bar{t}g$ scattering \label{sec:proj}}

In this section we discuss the construction of the helicity amplitudes for $gg\to t\bar{t}g$ and our computational framework.
We adopt the massive spinor helicity formalism of Ref.~\cite{Kleiss:1985yh}, where each massive momentum $p$ ($p^2 = m^2$) is first decomposed into two massless ones, as
\begin{equation}
	p^{\mu} = p^{\flat,\mu} + \frac{m^2}{2 \, p^{\flat} \cdot n} \, n^{\mu}\,,
\end{equation}
where $(p^\flat)^2 = 0$ and $n^2 = 0$. Massive spinors are then constructed from the massless reference vector $n$ via
\begin{equation}
	\label{eq:massivespinor}
	\bar{u}_+(p,m) = \frac{\la n | (\slashed{p}+m)}{\spA{n}{p^\flat}}, \qquad\qquad
	v_+(p,m)       = \frac{(\slashed{p}-m) | n \ra}{\spA{p^\flat}{n}} \,.
\end{equation}
We only need to consider the $+$ helicity configuration, since the $-$ one can be obtained from
the latter by exchanging the two massless momenta $p^\flat$ and $n$,
\begin{equation}
	\bar{u}_-(p,m) = \frac{\spA{p^\flat}{n}}{m} \bigg( \bar{u}_+(p,m)\bigg|_{p^\flat \leftrightarrow n} \bigg) \,.
\end{equation}
We refer to Ref.~\cite{Badger:2021owl} for further details on this formalism.

In order to derive helicity amplitudes which depend on arbitrary reference momenta for the top and anti-top quarks ($n_1$ and $n_2$, respectively),
we perform a decomposition using a basis made out of the $n_1$ and $n_2$ spinors~\cite{Badger:2021owl,Badger:2022mrb}:
\begin{align}
\label{eq:spindecomposition}
\begin{aligned}
        A^{(L)}(1^+_{\bar{t}},2^+_t,3^{h_3}_g,4^{h_4}_g,5^{h_5}_g;n_1,n_2) & 
= \frac{m_t \, \Phi^{h_3 h_4 h_5}}{\spA{1^{\flat}}{n_1}\spA{2^{\flat}}{n_2}}
  \bigg\lbrace  \spA{n_1}{n_2} s_{34} \; A^{(L),[1]}(1^+_{\bar{t}},2^+_t,3^{h_3}_g,4^{h_4}_g,5^{h_5}_g)  \\
  & \qquad + \spA{n_1}{3}\spA{n_2}{4} \spB{3}{4} \; A^{(L),[2]}(1^+_{\bar{t}},2^+_t,3^{h_3}_g,4^{h_4}_g,5^{h_5}_g) \\
  & \qquad + \spA{n_1}{3}\spA{n_2}{3} \frac{\spBB{3}{4|5}{3}}{s_{34}} \;  A^{(L),[3]}(1^+_{\bar{t}},2^+_t,3^{h_3}_g,4^{h_4}_g,5^{h_5}_g) \\
  & \qquad + \spA{n_1}{4}\spA{n_2}{4} \frac{\spBB{4}{5|3}{4}}{s_{34}} \;  A^{(L),[4]}(1^+_{\bar{t}},2^+_t,3^{h_3}_g,4^{h_4}_g,5^{h_5}_g) \bigg\rbrace \,,
\end{aligned}
\end{align}
where $s_{ij} = (p_i+p_j)^2$. 
The decomposition above applies to the $L$-loop bare, mass-renormalised and renormalised partial amplitudes ($A^{(L)}$, $A^{(L)}_{\mren}$ and $A^{(L)}_{\rm ren}$), as well as to the partial finite remainders ($R^{(L)}$).
The prefactors appearing in the ($n_1$,$n_2$) decomposition are chosen in such a way that the sub-amplitudes $A^{(L),[i]}$ are free of spinor phases and dimensionless.
We choose the helicity-dependent gluonic phase factors for the three independent gluon helicity configurations: ($+++$, $++-$, $+-+$)~as
\begin{equation}
\Phi^{+++} = \frac{\spB{3}{5}}{\spA{3}{4}\spA{4}{5}}\,, \qquad\quad
\Phi^{++-} = \frac{\spAA{5}{3|4}{5}}{\spA{3}{4}^2}\,, \qquad\quad
\Phi^{+-+} = \frac{\spAA{4}{5|3}{4}}{\spA{3}{5}^2}\,. \qquad\quad
\label{eq:gluonphase}
\end{equation}
We then evaluate the helicity amplitude at 4 different values of the ($n_1$,$n_2$) pair to obtain the sub-amplitudes $A^{(L),[i]}$ by inverting the following system of equations,
\begin{align}
	\label{eq:projectedhelamp}
\begin{aligned}
	A^{(L)}(1^+_{\bar{t}},2^+_t,3^{h_3}_g,4^{h_4}_g,5^{h_5}_g;p_3,p_3) & = \frac{m_t \Phi^{h_3 h_4 h_5}}{\spA{1^{\flat}}{3}\spA{2^{\flat}}{3}} 
	                                                                       \spA{3}{4}^2 \frac{\spBB{4}{5|3}{4}}{s_{34}} \;  A^{(L),[4]}(1^+_{\bar{t}},2^+_t,3^{h_3}_g,4^{h_4}_g,5^{h_5}_g)\,, \\
	A^{(L)}(1^+_{\bar{t}},2^+_t,3^{h_3}_g,4^{h_4}_g,5^{h_5}_g;p_3,p_4) & = \frac{m_t \Phi^{h_3 h_4 h_5}}{\spA{1^{\flat}}{3}\spA{2^{\flat}}{4}}		
	                                                                        \spA{3}{4} s_{34} \; A^{(L),[1]}(1^+_{\bar{t}},2^+_t,3^{h_3}_g,4^{h_4}_g,5^{h_5}_g) \,, \\
	A^{(L)}(1^+_{\bar{t}},2^+_t,3^{h_3}_g,4^{h_4}_g,5^{h_5}_g;p_4,p_3) & = - \frac{m_t \Phi^{h_3 h_4 h_5}}{\spA{1^{\flat}}{4}\spA{2^{\flat}}{3}}
									         \bigg[ \spA{3}{4} s_{34} \; A^{(L),[1]}(1^+_{\bar{t}},2^+_t,3^{h_3}_g,4^{h_4}_g,5^{h_5}_g)  \\
									   & \qquad \qquad + \spA{3}{4}^2 \spB{3}{4} \; A^{(L),[2]}(1^+_{\bar{t}},2^+_t,3^{h_3}_g,4^{h_4}_g,5^{h_5}_g) \bigg] \,, \\
	A^{(L)}(1^+_{\bar{t}},2^+_t,3^{h_3}_g,4^{h_4}_g,5^{h_5}_g;p_4,p_4) & = \frac{m_t \Phi^{h_3 h_4 h_5}}{\spA{1^{\flat}}{4}\spA{2^{\flat}}{4}}
                                                                               \spA{3}{4}^2 \frac{\spBB{3}{4|5}{3}}{s_{34}} \;  A^{(L),[3]}(1^+_{\bar{t}},2^+_t,3^{h_3}_g,4^{h_4}_g,5^{h_5}_g) \,.
\end{aligned}
\end{align}

We adopt the four-dimensional projection technique~\cite{Peraro:2019cjj,Peraro:2020sfm,Buccioni:2023okz} to compute the \textit{projected helicity amplitudes} appearing in the LHS of \cref{eq:projectedhelamp}.
We begin by writing the $gg\to t\bar{t}g$ amplitude in terms of four-dimensional tensor structures $\cT_i$ and form factors $\cF_i^{(L)}$,
\begin{equation}
        A^{(L)} = \sum_{i=1}^{32} \cT_i \, \cF_i^{(L)} \;,
        \label{eq:tensordecomposition}
\end{equation}
where
\begin{align}
\label{eq:Ttensor}
\begin{aligned}
	\cT_{1 \dots 8 }   &= m_t^2 \; \bar{u}(p_2) v(p_1) \; \Gamma_{1 \dots 8} \,, \\
	\cT_{9 \dots 16 }  &= m_t \; \bar{u}(p_2) \slashed{p}_3 v(p_1) \; \Gamma_{1 \dots 8} \,, \\
	\cT_{17 \dots 24 } &= m_t \; \bar{u}(p_2) \slashed{p}_4 v(p_1) \; \Gamma_{1 \dots 8} \,, \\
	\cT_{25 \dots 32 } &= \bar{u}(p_2)  \slashed{p}_3  \slashed{p}_4 v(p_1) \; \Gamma_{1 \dots 8} \,,  
\end{aligned}
\end{align}
with the following the tensor structures for the gluons,
\begin{align}
\label{eq:Gammatensor}
\begin{aligned}
	\Gamma_{1 } &=\epsv(p_3,q_3)  \cdot p_1 \; \epsv(p_4,q_4) \cdot p_1  \; \epsv(p_5,q_5) \cdot p_1 \,,  \\
	\Gamma_{2 } &=\epsv(p_3,q_3)  \cdot p_1 \; \epsv(p_4,q_4) \cdot p_1  \; \epsv(p_5,q_5) \cdot p_2 \,,  \\
	\Gamma_{3 } &=\epsv(p_3,q_3)  \cdot p_1 \; \epsv(p_4,q_4) \cdot p_2  \; \epsv(p_5,q_5) \cdot p_1 \,,  \\
	\Gamma_{4 } &=\epsv(p_3,q_3)  \cdot p_2 \; \epsv(p_4,q_4) \cdot p_1  \; \epsv(p_5,q_5) \cdot p_1 \,,  \\
	\Gamma_{5 } &=\epsv(p_3,q_3)  \cdot p_1 \; \epsv(p_4,q_4) \cdot p_2  \; \epsv(p_5,q_5) \cdot p_2 \,,  \\
	\Gamma_{6 } &=\epsv(p_3,q_3)  \cdot p_2 \; \epsv(p_4,q_4) \cdot p_1  \; \epsv(p_5,q_5) \cdot p_2 \,,  \\
	\Gamma_{7 } &=\epsv(p_3,q_3)  \cdot p_2 \; \epsv(p_4,q_4) \cdot p_2  \; \epsv(p_5,q_5) \cdot p_1 \,,  \\
	\Gamma_{8 } &=\epsv(p_3,q_3)  \cdot p_2 \; \epsv(p_4,q_4) \cdot p_2  \; \epsv(p_5,q_5) \cdot p_2 \,,
\end{aligned}
\end{align}
where $q_3$, $q_4$ and $q_5$ are the reference momenta for the gluon polarisation vectors, which we choose $q_3=p_4$, $q_4=p_3$ and $q_5=p_3$.
In \cref{eq:tensordecomposition,eq:Ttensor,eq:Gammatensor} we have dropped the arguments for the sake of readability.
The form factors $\cF^{(L)}_i$ can then be obtained by
\begin{equation}
	\cF_i^{(L)} = \sum_{j=1}^{32} \left( \Theta^{-1} \right)_{ij} \,\sum_{\mathrm{pol.}} \cT_j^\dagger A^{(L)} \;.
        \label{eq:formfactorsolution}
\end{equation}
where $\Theta_{ij} = \cT^\dagger_i \cT_j$ and we use following polarisation-vector sum,
\begin{equation}
	\sum_{\mathrm{pol.}} \vareps^{\mu *}(p,q) \, \vareps^\nu(p,q) = -g^{\mu\nu} + \frac{p^\mu q^\nu + q^\mu p^\nu}{p \cdot q}\,.
\end{equation}

To construct the helicity amplitudes we specify the helicity states of the gluons, top and anti-top quarks in the tensor structures $\cT_i$, together with their dependence on the reference vectors,
\begin{equation}
        A^{(L)}(1^+_{\bar{t}},2^+_t,3^{h_3}_g,4^{h_4}_g,5^{h_5}_g;n_1,n_2) = \sum_{i=1}^{32} \cT_i^{++h_3 h_4 h_5}(n_1,n_2) \; \cF_i^{(L)} \;,
        \label{eq:helicityamptensor}
\end{equation}
where
\begin{align}
\begin{aligned}
	T_{1 \dots 8 }^{++h_3 h_4 h_5}(n_1,n_2)   &= m_t^2 \; \bar{u}_+(p_2,n_2) \, v_+(p_1,n_1) \; \Gamma_{1 \dots 8}^{h_3 h_4 h_5} \,, \\
	T_{9 \dots 16 }^{++h_3 h_4 h_5}(n_1,n_2)  &= m_t \; \bar{u}_+(p_2,n_2) \slashed{p}_3 v_+(p_1,n_1) \; \Gamma_{1 \dots 8}^{h_3 h_4 h_5} \,,  \\
	T_{17 \dots 24 }^{++h_3 h_4 h_5}(n_1,n_2) &= m_t \; \bar{u}_+(p_2,n_2) \slashed{p}_4 v_+(p_1,n_1) \; \Gamma_{1 \dots 8}^{h_3 h_4 h_5} \,,  \\
	T_{25 \dots 32 }^{++h_3 h_4 h_5}(n_1,n_2) &= \bar{u}_+(p_2,n_2)  \slashed{p}_3  \slashed{p}_4 v_+(p_1,n_1) \; \Gamma_{1 \dots 8}^{h_3 h_4 h_5} \,, 
\end{aligned}
\end{align}
and
\begin{align}
\begin{aligned}
	\Gamma_{1 }^{h_3 h_4 h_5} &=\epsv(h_3,p_3,q_3)  \cdot p_1 \; \epsv(h_4,p_4,q_4) \cdot p_1  \; \epsv(h_5,p_5,q_5) \cdot p_1 \,,  \\
	\Gamma_{2 }^{h_3 h_4 h_5} &=\epsv(h_3,p_3,q_3)  \cdot p_1 \; \epsv(h_4,p_4,q_4) \cdot p_1  \; \epsv(h_5,p_5,q_5) \cdot p_2 \,,  \\
	\Gamma_{3 }^{h_3 h_4 h_5} &=\epsv(h_3,p_3,q_3)  \cdot p_1 \; \epsv(h_4,p_4,q_4) \cdot p_2  \; \epsv(h_5,p_5,q_5) \cdot p_1 \,,  \\
	\Gamma_{4 }^{h_3 h_4 h_5} &=\epsv(h_3,p_3,q_3)  \cdot p_2 \; \epsv(h_4,p_4,q_4) \cdot p_1  \; \epsv(h_5,p_5,q_5) \cdot p_1 \,,  \\
	\Gamma_{5 }^{h_3 h_4 h_5} &=\epsv(h_3,p_3,q_3)  \cdot p_1 \; \epsv(h_4,p_4,q_4) \cdot p_2  \; \epsv(h_5,p_5,q_5) \cdot p_2 \,,  \\
	\Gamma_{6 }^{h_3 h_4 h_5} &=\epsv(h_3,p_3,q_3)  \cdot p_2 \; \epsv(h_4,p_4,q_4) \cdot p_1  \; \epsv(h_5,p_5,q_5) \cdot p_2 \,,  \\
	\Gamma_{7 }^{h_3 h_4 h_5} &=\epsv(h_3,p_3,q_3)  \cdot p_2 \; \epsv(h_4,p_4,q_4) \cdot p_2  \; \epsv(h_5,p_5,q_5) \cdot p_1 \,,  \\
	\Gamma_{8 }^{h_3 h_4 h_5} &=\epsv(h_3,p_3,q_3)  \cdot p_2 \; \epsv(h_4,p_4,q_4) \cdot p_2  \; \epsv(h_5,p_5,q_5) \cdot p_2 \,.
\end{aligned}        %
\end{align}
Finally, the helicity amplitudes are derived from the form factors by combining \cref{eq:helicityamptensor,eq:formfactorsolution}, as
\begin{equation}
        A^{(L)}(1^+_{\bar{t}},2^+_t,3^{h_3}_g,4^{h_4}_g,5^{h_5}_g;n_1,n_2) = \sum_{i,j=1}^{32} T_i^{++h_3 h_4 h_5}(n_1,n_2) \,
	 \left( \Theta^{-1} \right)_{ij} \,\sum_{\mathrm{pol.}} T_j^\dagger A^{(L)}  \;.
        \label{eq:helicityamplitude}
\end{equation}
The projected helicity amplitudes on the LHS of \cref{eq:projectedhelamp} are obtained by evaluating \cref{eq:helicityamplitude} at the corresponding value of the ($n_1$,$n_2$) pair.

\smallskip

\begin{figure}[t!]
    \centering
    \begin{subfigure}[b]{0.3\textwidth}
        \centering
        \includegraphics[height=3.5cm]{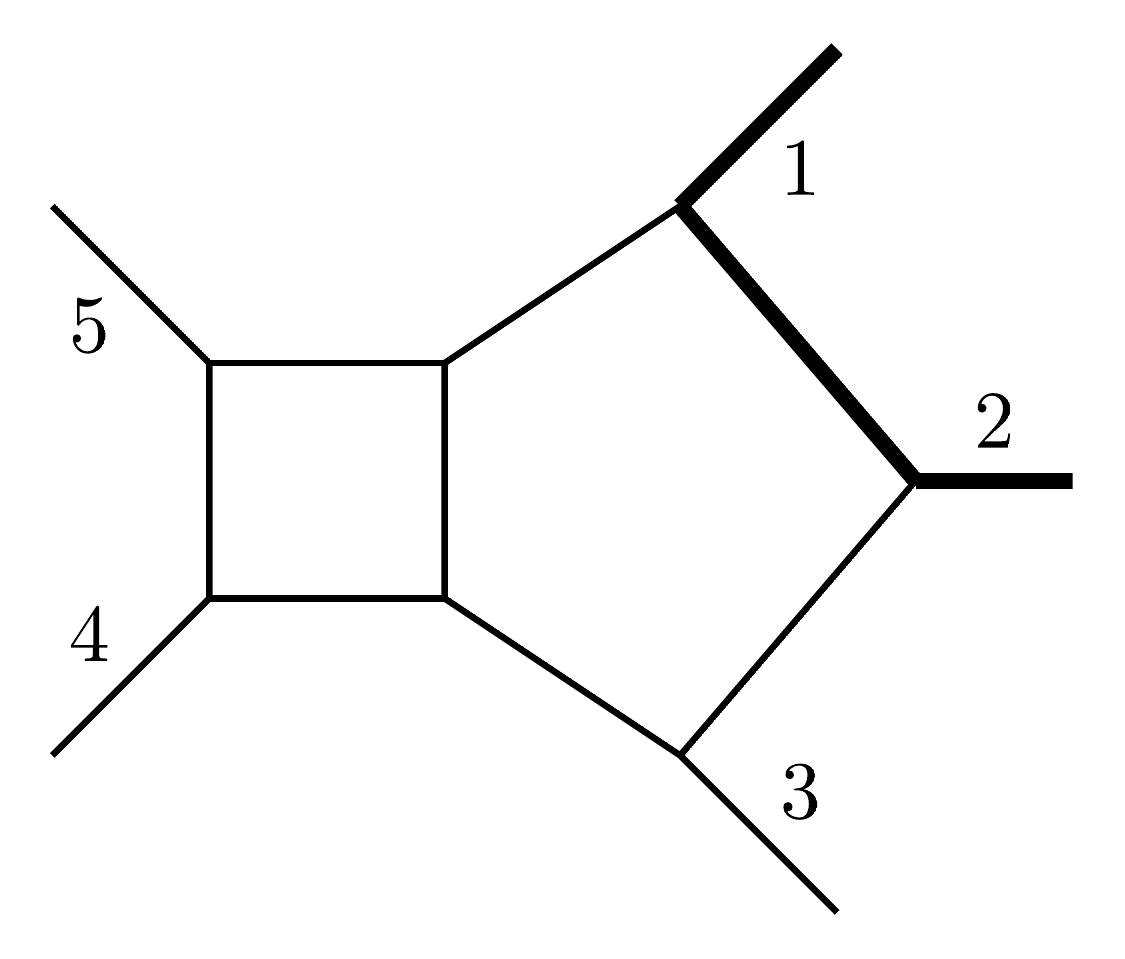}
        \caption{Family ${\rm PB}_A$.}
        \label{fig:PBA}
    \end{subfigure}
    \hfill
    \begin{subfigure}[b]{0.3\textwidth}
        \centering
        \includegraphics[height=3.5cm]{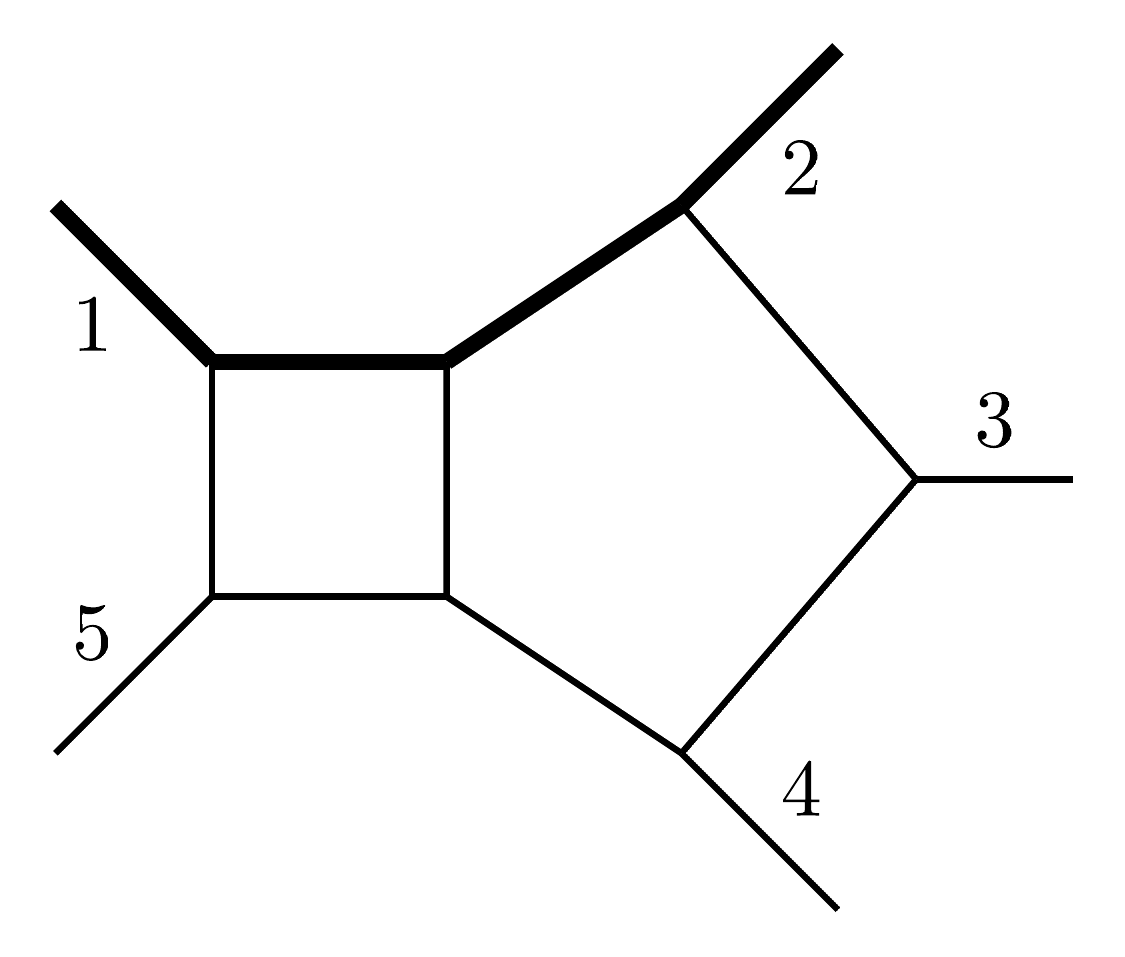}
        \caption{Family ${\rm PB}_B$.}
        \label{fig:PBB}
    \end{subfigure}
    \hfill
    \begin{subfigure}[b]{0.3\textwidth}
        \centering
        \includegraphics[height=3.5cm]{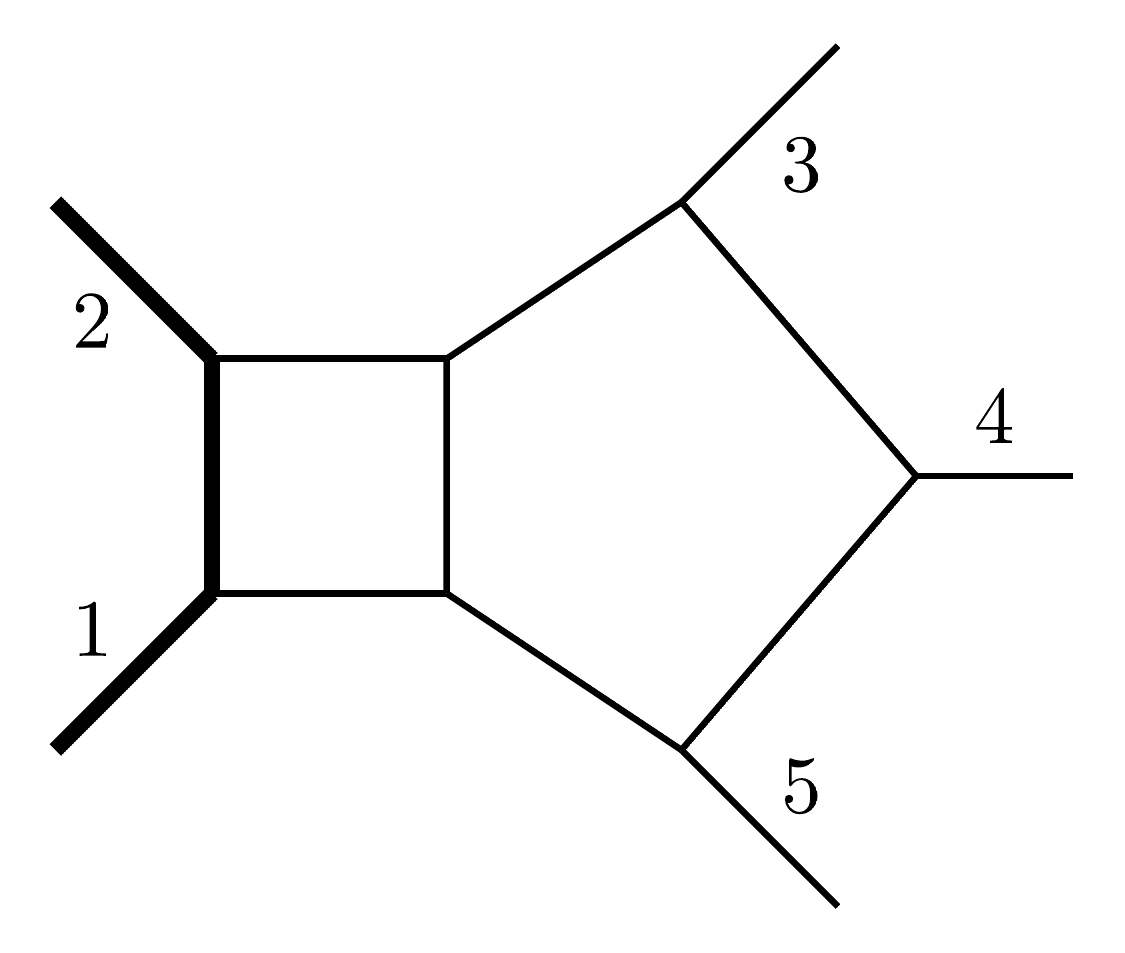}
        \caption{Family ${\rm PB}_C$.}
        \label{fig:PBC}
    \end{subfigure}
    
    \vspace{1em} 

    \begin{subfigure}[b]{0.3\textwidth}
        \centering
        \includegraphics[height=3.75cm]{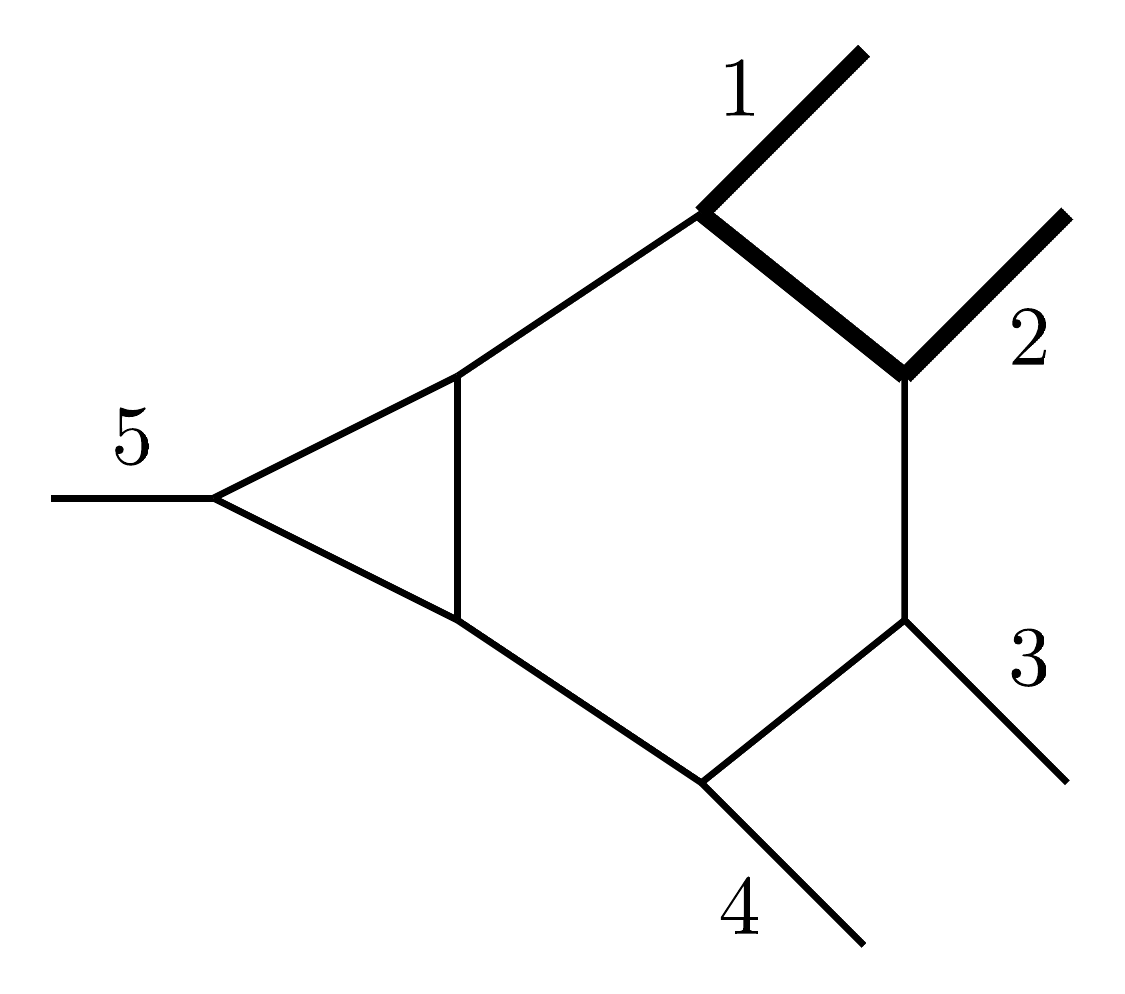}
        \caption{Family ${\rm HT}_A$.}
        \label{fig:HTA}
    \end{subfigure}
    \hfill
    \begin{subfigure}[b]{0.3\textwidth}
        \centering
        \includegraphics[height=3.75cm]{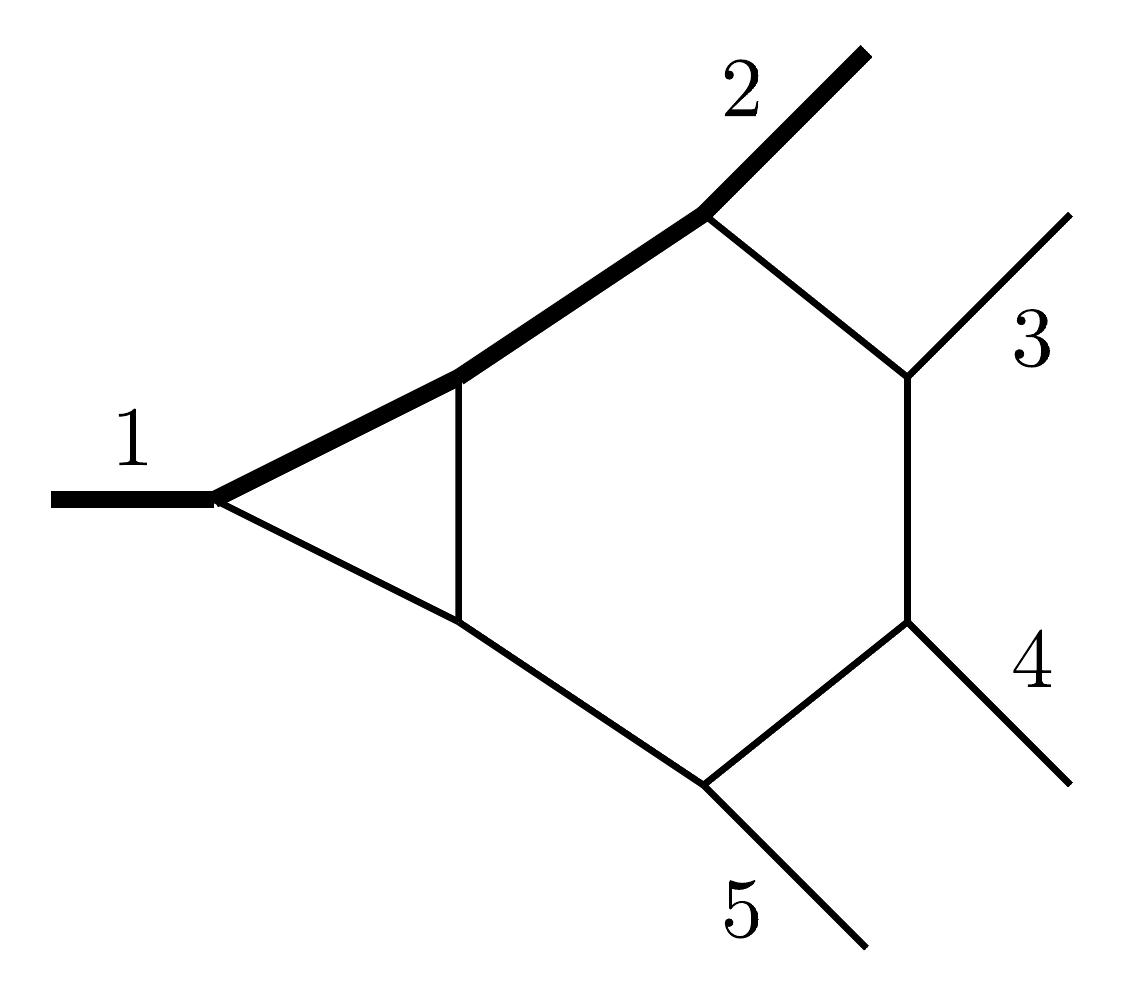}
        \caption{Family ${\rm HT}_B$.}
        \label{fig:HTB}
    \end{subfigure}
    \hfill
    \begin{subfigure}[b]{0.3\textwidth}
        \centering
        \includegraphics[height=3.75cm]{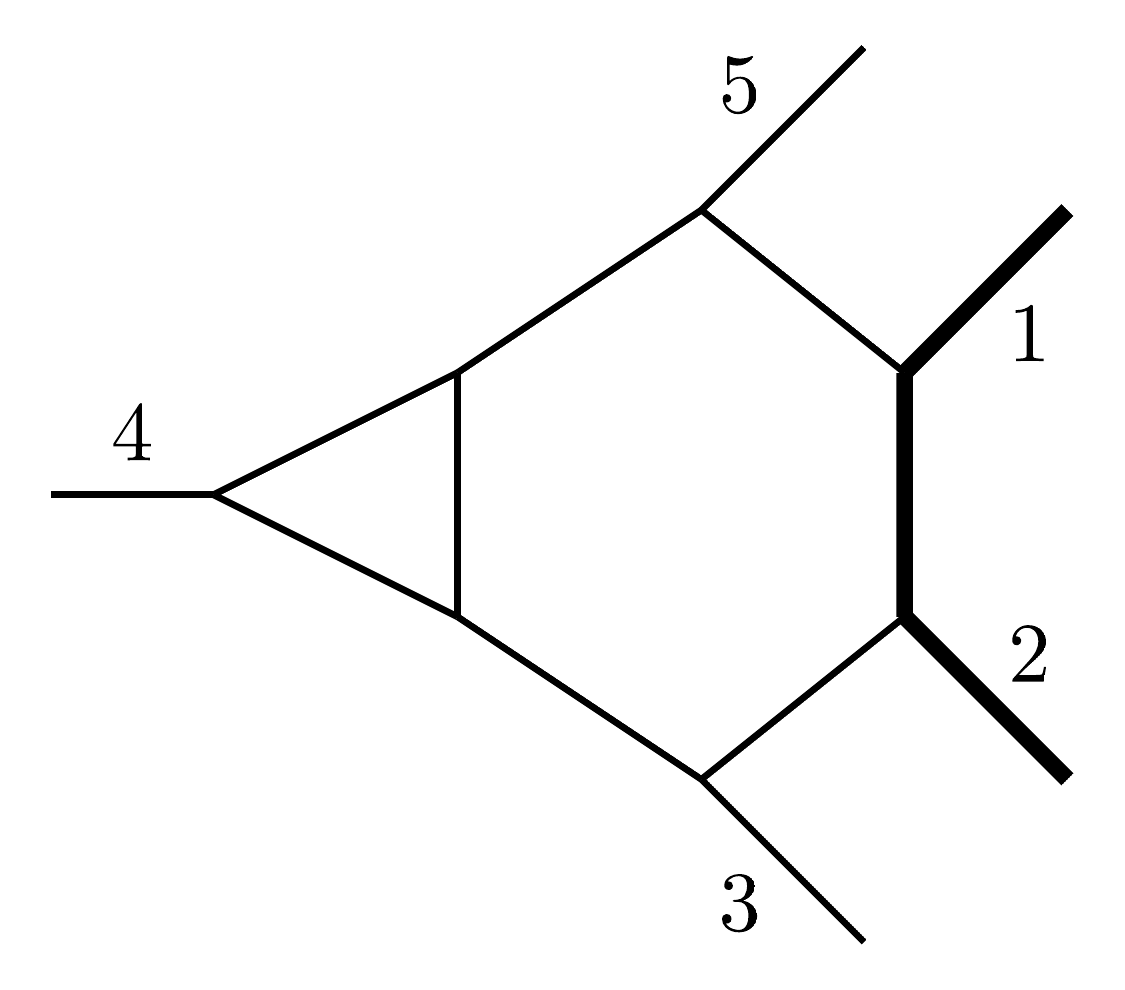}
        \caption{Family ${\rm HT}_C$.}
        \label{fig:HTC}
    \end{subfigure}

	\caption{Integral families relevant for the leading-colour two-loop $gg\rightarrow t\bar{t}g$ amplitude. 
	The hexagon-triangle (${\rm HT}$) families, in the second row, can be reduced to pentagon-box (${\rm PB}$) families, in the first row, except for 4 master integrals of ${\rm HT}_B$ which can instead be expressed as products of one-loop integrals. 
	Four more families, $\mathrm{PB}_{A}'$, $\mathrm{PB}_{B}'$, $\mathrm{HT}_{A}'$ and $\mathrm{HT}_{B}'$, are obtained by exchanging $1 \leftrightarrow 2$ and $3 \leftrightarrow 5$ in $\mathrm{PB}_{A}$, $\mathrm{PB}_{B}$, $\mathrm{HT}_{A}$ and $\mathrm{HT}_{B}$, respectively.}
    \label{fig:topologies}
\end{figure}

We now move on to the discussing how we evaluate numerically the projected helicity amplitudes.
We apply the finite-field framework already used in several two-loop five-point computations
(see for example Refs.~\cite{Badger:2024sqv,Badger:2024awe} and references therein
for the details). Here we content ourselves with outlining the key steps.
We first construct the \textit{contracted amplitude} $\sum_{\mathrm{pol.}} T_i^\dagger A^{(L)}$ by generating the Feynman diagrams with \textsc{Qgraf}~\cite{Nogueira:1991ex}
and contracting them with the conjugated tensor structures in \cref{eq:Ttensor}. The loop-integral topologies associated with the Feynman diagrams are mapped onto
a set of integral families with the highest number of propagators (8 for five particles at two loops).
Algebraic manipulations and simplifications of the contracted amplitudes are carried out using a collection of \textsc{Mathematica} and \textsc{Form}~\cite{Ruijl:2017dtg} scripts.
As a result, the contracted amplitude
$\sum_{\mathrm{pol.}} T_i^\dagger A^{(L)}$ is expressed as a linear combination of scalar
Feynman integrals, which are then reduced to the set of master integrals of Ref.~\cite{Badger:2024fgb} by means of integration-by-part (IBP) reduction~\cite{Tkachov:1981wb,Chetyrkin:1981qh,Laporta:2000dsw}. 
The planar two-loop integral families entering the IBP reduction are shown in \cref{fig:topologies}, and comprise five pentagon-boxes (PB) and five hexagon-triangles (HT).
All the scalar Feynman integrals belonging to the HT families are reduced to master integrals in the PB families except for a subset of master integrals in the ${\rm HT}_B$ and ${\rm HT}_B'$ families that factorise into products of one-loop integrals.
Before carrying out IBP reduction, we add to the bare contracted amplitude the contributions from the mass-renormalisation counterterms according to \cref{eq:Amren}.
In order to do this, we write the mass counterterms in \cref{eq:Amren} 
in terms of scalar Feynman integrals using the same convention as in the bare amplitude.
From this stage on we therefore work with the mass-renormalised amplitude rather than its bare counterpart, with the advantage of gauge invariance.

With respect to Ref.~\cite{Badger:2024fgb} and \cref{sec:spfunc_construction}, we remove from the master integrals the overall square-root normalisations.
The next step is the Laurent expansion around $\eps=0$.
For the rational coefficients of the master integrals this is done within the finite-field framework.
The master integrals are instead expressed order by order in $\eps$ as polynomials in the set of special functions and transcendental constants defined in \cref{sec:special_functions}, as well as square roots originating from the different normalisation.
We denote cumulatively the special functions and transcendental constants by $F$ and the nine square roots by $\sqrtsymb$.
As a result, the mass-renormalised contracted amplitude takes the form
\begin{equation} \label{eq:contractedamplitude}
	\sum_{\mathrm{pol.}} T_i^\dagger A^{(L)}_{\mren} = \sum_{s=-2L}^{4 - 2 L} \sum_j \eps^s \; c_{ij,s}\bigl(\vec{d} \; \bigr) \; \mathrm{mon}_j\left(F,\sqrtsymb \right)  + \mathcal{O}\left(\eps^{5 - 2 L}\right)\,,
\end{equation} 
where $\mathrm{mon}_j$ denotes monomials of special functions
$F$ and square roots $\sqrtsymb$, and we recall that $\vec{d}$ is the set of scalar invariants defined in \cref{eq:dijmtset}.
The finite remainder is obtained by subtracting the IR and remaining UV poles from the mass-renormalised amplitude as prescribed in \cref{eq:UVRen,eq:IRRen}.
To this end, we first express the UV and IR subtraction terms
in the same set of special functions as the two-loop amplitude. The subtraction terms are then Laurent expanded in $\eps$ through the same order
as the bare amplitude. The finite remainder of the contracted amplitude, $\sum_{\mathrm{pol.}} T_i^\dagger R^{(L)}$, is given by 
\begin{equation}
\label{eq:contractedfinrem}
\sum_{\mathrm{pol.}} T_i^\dagger R^{(L)} =  \sum_{s=0}^{4 - 2 L} \sum_j \eps^s \; 
	r_{ij,s}\bigl(\vec{d} \; \bigr) \; \mathrm{mon}_j\left(F,\sqrtsymb\right) + \cO\left(\eps^{5 - 2 L}\right) \,.
\end{equation} 
We recall that the finite remainder admits the same colour and spin-basis decomposition as the bare amplitude as specified in \cref{eq:ColDec,eq:spindecomposition}. 
We can further derive the helicity-dependent finite remainder $R^{(L)}$ in
terms of special functions following \cref{eq:helicityamplitude}, as
\begin{align}
\label{eq:helicityamplitudepfuncs0}
     R^{(L)}(1^+_{\bar{t}},2^+_t,& 3^{h_3}_g, 4^{h_4}_g,5^{h_5}_g;n_1,n_2)    \nonumber \\
& = \sum_{i,j=1}^{32} \sum_k \sum_{s=0}^{4 - 2 L} T_i^{++h_3 h_4 h_5}(n_1,n_2) 
	\left( \Theta^{-1} \right)_{ij} \; \eps^s \;r_{jk,s}\bigl(\vec{d}\; \bigr) \mathrm{mon}_k\left(F,\sqrtsymb\right) + \cO\left(\eps^{5 - 2 L}\right) \,, \\
\label{eq:helicityamplitudepfuncs}        
	& = \sum_i \sum_{s=0}^{4 - 2 L} \eps^s \; \hat{r}^{h_3 h_4 h_5}_{i,s}(\lbrace p \rbrace,n_1,n_2) \; \mathrm{mon}_i\left(F,\sqrtsymb\right) + \cO\left(\eps^{5 - 2 L}\right) \,,
\end{align}
where $\hat{r}^{h_3 h_4 h_5}_{i,s}(\lbrace p \rbrace,n_1,n_2 )$ is made of spinor brackets and scalar products involving external momenta $p_i$ and reference
vectors $n_i$. In the case of the projected helicity finite remainder (cf.~\cref{eq:projectedhelamp}), we have
\begin{align}
     R^{(L)}(1^+_{\bar{t}},2^+_t,3^{h_3}_g,4^{h_4}_g,5^{h_5}_g;p_a,p_b) & = 
     \sum_i \sum_{s=0}^{4 - 2 L} \eps^s \; \tilde{r}^{h_3 h_4 h_5}_{i,s;ab}(\lbrace p \rbrace) \; \mathrm{mon}_i\left(F,\sqrtsymb\right) + \cO\left(\eps^{5 - 2 L}\right) \,,
\label{eq:projectedfinrempfuncs}        
\end{align}
where $( p_a,p_b ) \in \lbrace (p_3,p_3), (p_3,p_4), (p_4,p_3), (p_4,p_4) \rbrace$
and the rational coefficients $\tilde{r}^{h_3 h_4 h_5}_{i,s;ab}$ are now expressed fully in terms of on-shell $gg\to t\tb g$ kinematics.

Starting from the analytic expression of the contracted amplitudes $\sum_{\mathrm{pol.}} T_i^\dagger A^{(L)}$, the computation is organised as a dataflow graph where all rational operations among rational functions are performed numerically in a finite field within the framework \textsc{FiniteFlow}~\cite{Peraro:2016wsq,Peraro:2019svx}.
This includes in particular the IBP reduction to master integrals and the Laurent expansion of the rational coefficients.
We simplify the IBP-reduction step by generating optimised systems of IBP relations with \textsc{NeatIBP}~\cite{Wu:2023upw}.
The conversion of the contracted amplitudes into projected helicity amplitudes according to
\cref{eq:helicityamplitudepfuncs0} requires a rational parameterisation of the external
kinematics including the spinor brackets.
For this purpose we use the momentum-twistor parameterisation defined in  Ref.~\cite{Badger:2022mrb}, in terms of the variables
\begin{align} \label{eq:MTvars}
 \vec{x} = \bigl(s_{34}, t_{12}, t_{23}, t_{45}, t_{51}, x_{5123} \bigr)\,.
\end{align}

In this work, we obtain numerical results for the $gg\to t\bar{t}g$ helicity amplitude.
In other words, we do not compute the analytic expression of the rational coefficients of the special function monomials, but obtain numerical values for them at sample phase-space points in the physical region.
We set $\vec{d}$ at the chosen phase-space point and rationalise it, and carry out the computation from the unreduced mass-renormalised contracted amplitude, $\sum_{\mathrm{pol.}} T_i^\dagger A^{(L)}_{\mren}$, to
obtain the corresponding representation in terms of special functions  (cf.~\cref{eq:contractedamplitude}), numerically over finite fields.
By putting together the values of the coefficients of the special-function monomials in sufficiently many prime fields, we can perform a rational reconstruction and obtain their values in $\mathbb{Q}$.
The subsequent set of operations are done outside of the finite-field framework.
We subtract the UV and IR poles from the mass-renormalised contracted amplitude to derive the finite remainder representation as in~\cref{eq:contractedfinrem}, where the special functions are kept symbolic
while the corresponding rational coefficients are evaluated numerically at the chosen phase-space point, and carry out the conversion to the projected helicity finite remainder
in \cref{eq:projectedfinrempfuncs} via momentum-twistor variables.\footnote{One of the variables in the momentum twistor parameterisation, $x_{5123}$, 
has a non-zero imaginary part in the physical phase space. While this is not an issue in our numerical computation since we perform the conversion to the helicity amplitude outside of the finite-field framework, 
carrying out such a conversion within the finite-field approach requires a modification of the standard rational reconstruction algorithm to accommodate complex numbers.}
As a result, we obtain numerical values of the coefficients of the special-function monomials, i.e., $\tilde{r}^{h_3 h_4 h_5}_{i,s;ab}(\lbrace p \rbrace)$
in \cref{eq:projectedfinrempfuncs}, for the three independent helicity configurations.
The evaluation of the finite remainders is then completed by the evaluation of the special functions, which is discussed in \cref{sec:spfunc_evaluation,sec:performance}.


\section{Laurent expansion of the master integrals}
\label{sec:special_functions}

In this section we discuss our method to express the master integrals in terms of a set of special functions designed to accomplish three key goals: the analytic subtraction of the UV/IR poles of the amplitudes, the simplification of the finite remainders, and a more efficient numerical evaluation with respect to what was previously available in the literature~\cite{Badger:2022hno,Badger:2024fgb}.
The starting point is the method of the so-called \emph{pentagon functions}~\cite{Gehrmann:2018yef,Chicherin:2020oor,Badger:2021nhg,Chicherin:2021dyp,Abreu:2023rco}, which has proven extremely successful in the computation of two-loop five-particle amplitudes with no internal massive propagators (see also Refs.~\cite{Henn:2022ydo,Badger:2023xtl,AH:2023ewe,Henn:2024ngj,Gehrmann:2024tds,Liu:2024ont} for the application of similar methods to other cases).
This method applies to those cases where the basis $\vec{g}_{\rm F}$ of all relevant integral families $\mathrm{F}$ satisfies a differential equation (DE) in the canonical form~\cite{Henn:2013pwa}
\begin{align}
 \label{eq:DEcanonical}
 \d \vec{g}_{\rm F}(X; \eps) = \eps \, \sum_i A_i \, \d \log W_i(X) \cdot \vec{g}_{\rm F}(X; \eps) \,,
\end{align}
where $X$ denotes the kinematic variables (in this case $X=\vec{d}$, cf.~\cref{eq:dijmtset}), $\eps = (4-D)/2$ is the dimensional regulator, $\d$ is the total differential w.r.t.\ $X$, the $A_i$'s are constant matrices with entries in $\mathbb{Q}$, and the $W_i(X)$'s are algebraic functions of $X$ called letters.
The method allows us to write the solution to \cref{eq:DEcanonical} in terms of a \emph{basis} of special functions, i.e., of a set of algebraically independent special functions.

In this work, we initiate the extension of this approach to Feynman integrals which satisfy DEs in a form different from \cref{eq:DEcanonical}.
We explain our algorithm to construct a set of special functions to express the solution to non-canonical DEs in \cref{sec:spfunc_construction}, discuss how we can evaluate them numerically in \cref{sec:spfunc_evaluation}, 
and present benchmark evaluations in \cref{sec:performance}.

\begin{figure}[t]
    \centering
    \begin{subfigure}{0.32\textwidth}
        \centering
        \includegraphics[scale=0.16]{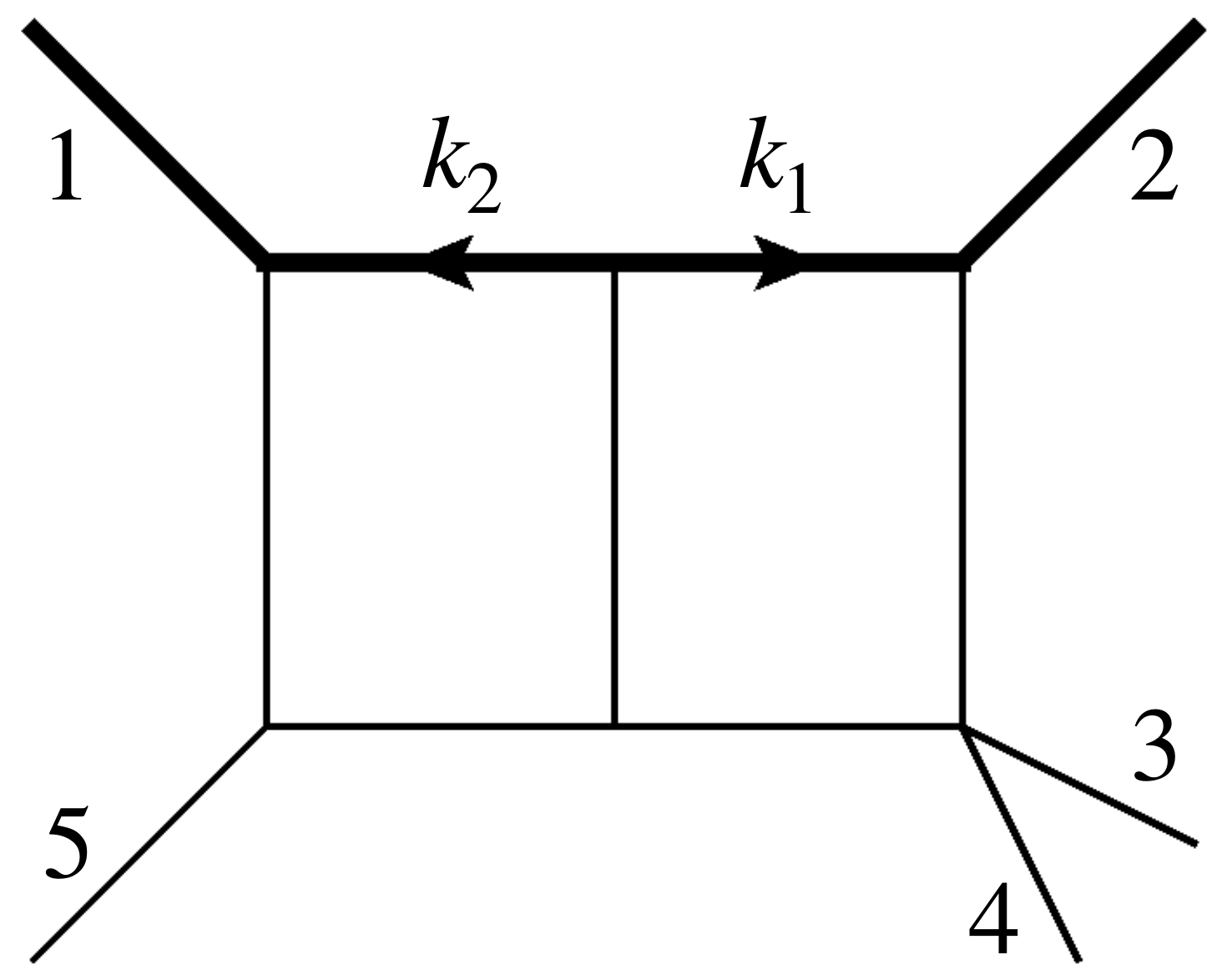}
        \caption{$g_{15}$}
        \label{fig:sector_15}
    \end{subfigure}
    \begin{subfigure}{0.32\textwidth}
        \centering
        \includegraphics[scale=0.16]{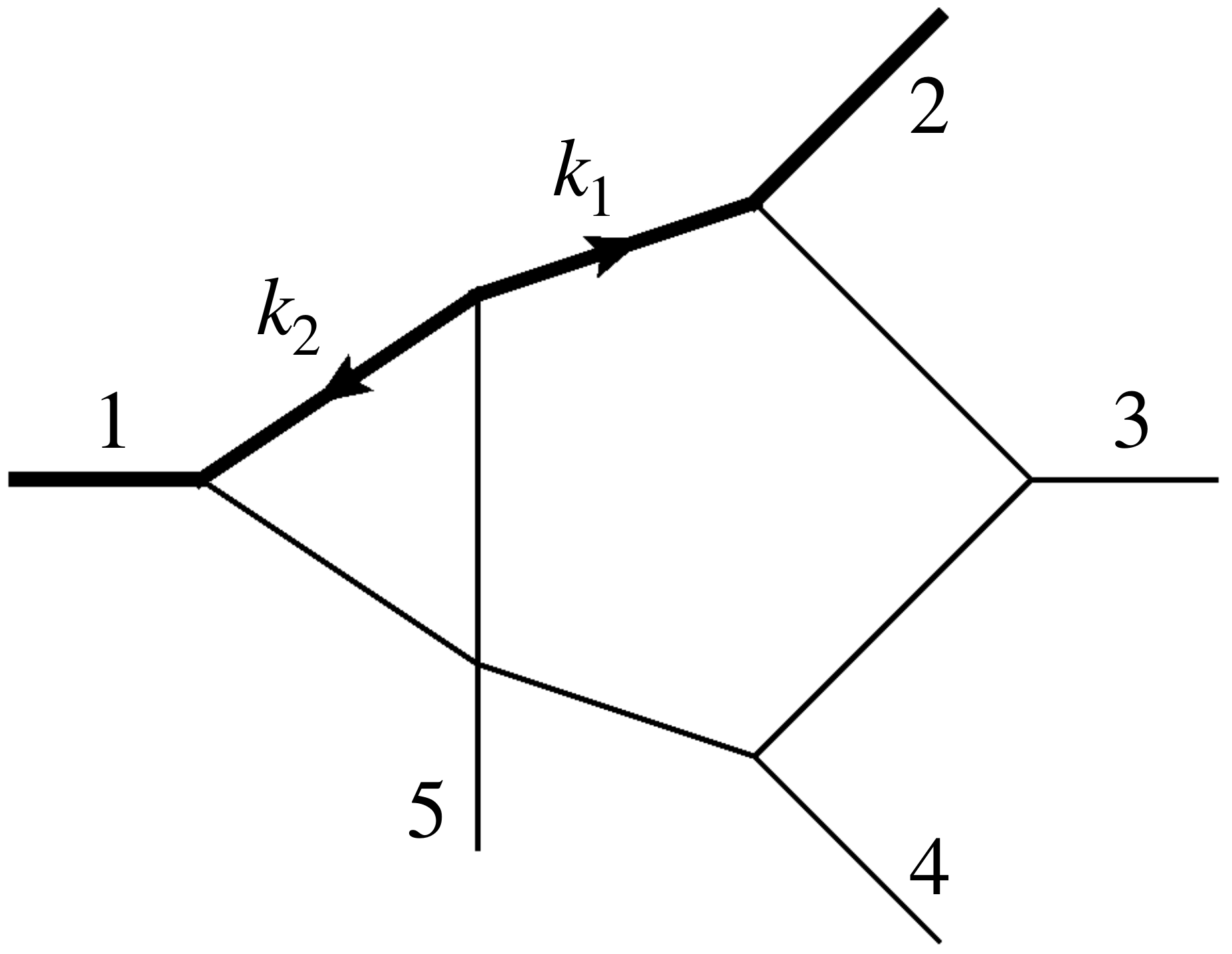}
        \caption{$g_{19}$, $g_{20}$}
        \label{fig:sector_nested}
    \end{subfigure}
\hfill
    \begin{subfigure}{0.32\textwidth}
        \centering
        \includegraphics[scale=0.16]{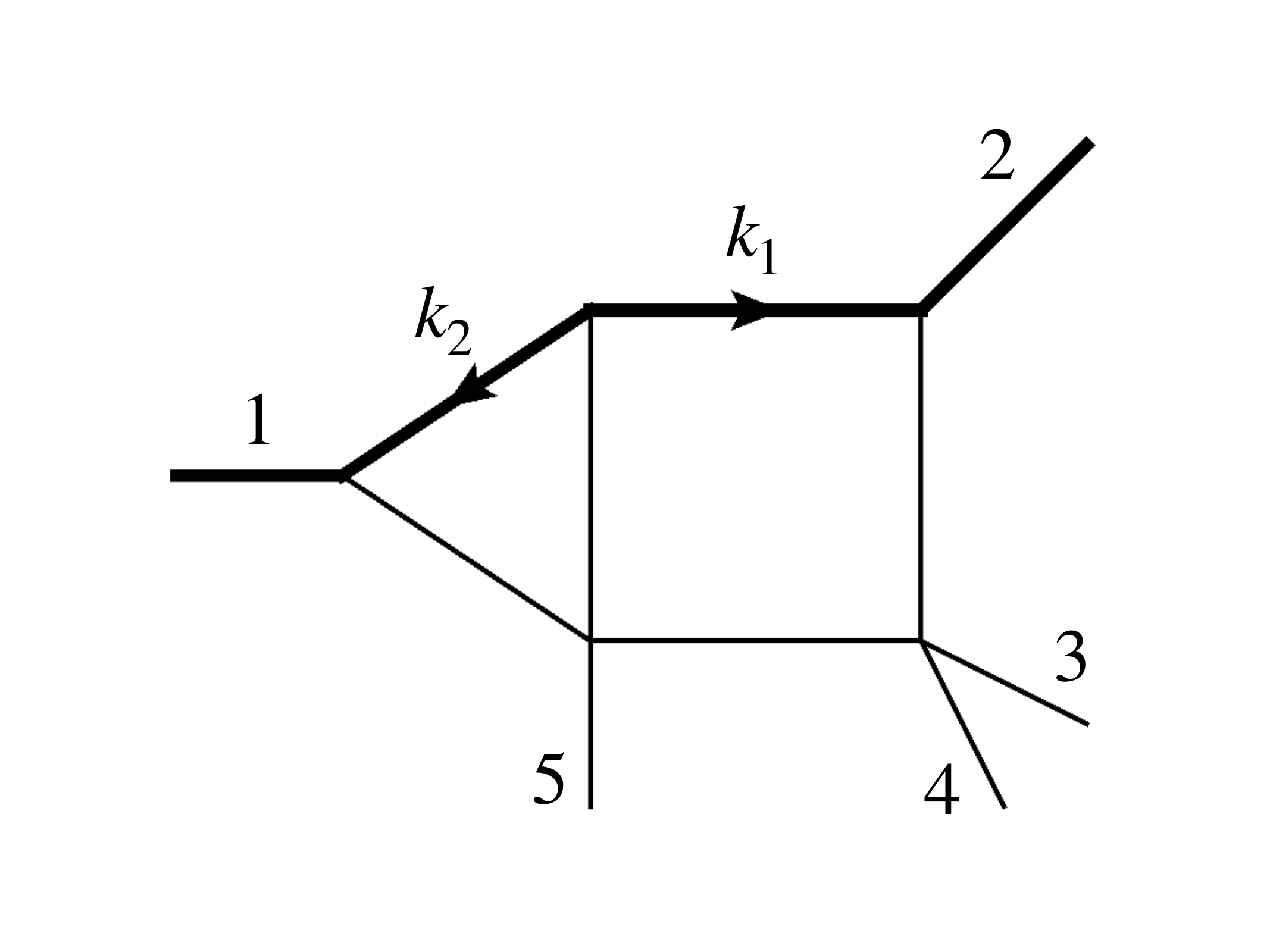}
        \caption{$g_{35}$, $g_{36}$, $g_{37}$}
        \label{fig:sector_elliptic}
    \end{subfigure}
    \caption{Graphs representing the sectors of the family ${\rm PB}_B$ which contain master integrals that we did not express in terms of iterated integrals at order $\eps^4$.
    The external momenta are all outgoing, the thick line denotes the top, and the arrows indicate the momentum's direction. 
    The sub-captions list the relevant master integrals.
    They are defined by multiplying the scalar propagators associated with the graphs by the following numerators under the integral sign:
    \begin{minipage}[t]{\linewidth}
    \vspace{-0.43cm}
     \begin{align*}
       \begin{aligned}
        & N_{15} = \eps^4 \, d_{15}^2 \, (k_2 + p_2)^2 + \ldots \,, \\
        & N_{19} = \eps^4 \, d_{45} \, \mathrm{tr}\bigl(\gamma_5 \slashed{p}_3 (\slashed{k}_1-\slashed{p}_2-\slashed{p}_3) \slashed{p}_4 \slashed{p}_2 \bigr) \,, \\
        & N_{20} = \eps^4 \, d_{45} \, \mathrm{tr}\bigl(\slashed{p}_3 (\slashed{k}_1-\slashed{p}_2-\slashed{p}_3) \slashed{p}_4 \slashed{p}_2 \bigr) \,,
        \end{aligned}
        \qquad
        \begin{aligned}
        & N_{35} =  \eps^4 \, d_{15} (d_{12} + m_t^2) \,, \\
        & N_{36} = 2 \, \eps^4 \, \sqrt{(d_{15}-d_{34})^2-2 d_{34} m_t^2} \, (k_1 \cdot p_1) \,, \\
        & N_{37} = 2 \, \eps^4 (2 \eps - 1) \, d_{15} \, (k_2 \cdot p_2) \,,
        \end{aligned}
     \end{align*}
     \vspace{0cm}
  \end{minipage}
  where the ellipsis denotes sub-sector terms.
  We recall that the sector in figure~(b) contains the nested square root, whereas the sector in figure~(c) involves elliptic functions.
}
    \label{fig:problematic_sectors}
\end{figure}

\subsection{Construction of a (over-complete) basis of special functions}
\label{sec:spfunc_construction}
We begin by quickly reviewing the method of Ref.~\cite{Abreu:2023rco}.
The inputs are canonical DEs for all the relevant one- and two-loop integral families ${\rm F}$, and numerical `boundary' values of all basis integrals $\vec{g}_{\rm F}$ at an arbitrary phase-space point $X_0$.
Using this information, one writes the solution to the canonical DEs in terms of Chen iterated integrals~\cite{Chen:1977oja} (see e.g.\ Ref.~\cite{Abreu:2022mfk} for a review), order by order in $\eps$,
\begin{align} \label{eq:epsexp}
 \vec{g}_{\rm F}(X;\eps) = \sum_{k\ge 0} \eps^k \, \vec{g}^{\, (k)}_{\rm F}(X) \,.
\end{align}
We truncate the expansion at the highest order required to compute two-loop amplitudes up to their finite part.
By reducing the amplitude to master integrals as discussed in \cref{sec:proj}, we know that this is $k=4$.\footnote{Higher orders in $\eps$ may in general be required if the master-integral basis is not canonical.}
The order in $\eps$ in the canonical case matches the transcendental weight, and
the weight-0 $\eps$-coefficients $\vec{g}_{\rm F}^{(0)}$ are constant and rational.
Next, we exploit the $\mathbb{Q}$-linear independence of iterated integrals and the shuffle product to select, out of all $\eps$-coefficients $\{\vec{g}_{\rm F}^{(k)}, \forall \, \mathrm{F}, i=1,2,3,4 \}$, a subset of elements which are algebraically independent.
We denote them by $\{F_i^{(k)}\}$, where $k$ denotes the transcendental weight.
This is achieved by Gaussian elimination, with an ordering which is chosen to impose certain useful criteria.
In particular, products of lower-weight $\eps$-coefficients are preferred over higher-weight ones, and one-loop $\eps$-coefficients over two-loop ones.
As a result, the number of higher-weight functions is minimised, and we know a priori that the $\eps$-poles of the amplitudes contain only the subset of functions coming from the one-loop integrals.
More criteria are imposed on a case-by-case basis, e.g.\ to make manifest the cancellation of certain letters from the amplitudes/finite remainders.
We perform this procedure first at the symbol level~\cite{Goncharov:2010jf}, i.e., by setting to zero all $\vec{g}^{(k)}_{\rm F}(X_0)$ for $k>0$.
The result is then lifted to iterated integral by ansatzing.
Finally, the outputs are:
\begin{itemize}
 \item a list of algebraically independent $\eps$-coefficients, $\{F_i^{(k)}\}$, dubbed pentagon functions,
 \item the expression of all $\eps$-coefficients $\vec{g}_{\rm F}^{(k)}(X)$ as polynomials in the ring generated by the pentagon functions, $\zeta_2$ and $\zeta_3$, with coefficients in $\mathbb{Q}$,
 \item polynomial relations among the values $\vec{g}_{\rm F}^{(k)}(X_0)$ with $k>0$, which are checked numerically.
\end{itemize}

The method summarised above applies straightforwardly to the one-loop family ($\mathrm{P}_A$) and to three of the five two-loop families relevant for $t\bar{t}j$-production at leading colour, namely $\mathrm{PB}_A$, $\mathrm{PB}_{A}'$ and $\mathrm{PB}_C$.\footnote{We do not need to consider the hexagon-triangle families here, as they can be expressed in terms of pentagon-box integrals and products of one-loop integrals.}
We recall that the superscript $'$ in the family name denotes the permutation $(1\to 2, 2 \to 1, 3\to 5, 4 \to 4, 5\to 3)$ of the external legs.
The canonical DEs for the permuted families are obtained by permuting the DEs for the families in the standard ordering derived in Refs.~\cite{Badger:2022hno,Badger:2024fgb}.
In doing this, we add 28 new letters and two new square roots (permutations of $\Lambda_1$ and $\Lambda_3$) to the alphabet of Ref.~\cite{Badger:2024fgb}.
The available DEs for the two remaining families (${\rm F} = \mathrm{PB}_{B}, \mathrm{PB}_{B}'$), instead, have the non-canonical form
\begin{align}
 \label{eq:DEnonCanonical}
 \d \vec{g}_{\rm F}(X; \eps) = \Omega_{\rm F}(X;\eps)  \cdot \vec{g}_{\rm F}(X; \eps) \,,
\end{align}
where the connection matrix $\Omega_{\rm F}(X;\eps)$ is a degree-2 polynomial in $\eps$,
\begin{align}
\Omega_{\rm F}(X;\eps) = \sum_{k=0}^2 \eps^k \, \Omega^{(k)}_{\rm F}(X) \,,
\end{align}
with
\begin{align}
 \label{eq:DEnonCanonicalMatrix}
\Omega_{\rm F}^{(k)}(X) = \sum_i A^{({\rm F})}_{k,i} \, \d \log W_i(X) + \sum_j  B^{({\rm F})}_{k,j} \, \omega_j(X) \,.
\end{align}
Here, $\omega_j(X)$ are $\mathbb{Q}$-linearly independent non-logarithmic one-forms, written in terms of differentials of the variables $X$ and algebraic coefficients.
In order to clarify what we mean here by ``non-logarithmic,'' we note that all but two one-forms---$\omega_{34}$ and $\omega_{49}$---are not closed, and thus have no primitive.
Of the closed one-forms, $\omega_{34}$ has an algebraic primitive, whereas for $\omega_{49}$ one can verify by integration that its primitive contains logarithms.
We did not try to express $\omega_{49}$ as a $\d \log$ because, from its position in the connection matrix, we expect its primitive to involve nested square roots, which are not supported by the public packages used for the numerical solution of the DEs (see \cref{sec:spfunc_evaluation}).
What matters for the following is that the one-forms $\omega_i(X)$ are $\mathbb{Q}$-linearly independent, and that the deviation from the canonical form is restricted to a small subset of MIs: $S=\{15,19,20,35,36,37\}$.
We recall their definition from Ref.~\cite{Badger:2024fgb} in \cref{fig:problematic_sectors}.
Explicitly, we have that
\begin{itemize}
\item the entry $(\Omega_{\rm F})_{ab}$ is canonical for $a,b\neq S$, i.e., it is $\eps$-factorised and involves $\d \log$'s only;
\item only the derivatives of the 37th MI are quadratic in $\eps$, i.e., $[A_{2,i}^{(\rm F)}]_{ab} = 0 = [B_{2,i}^{(\rm F)}]_{ab}$ for $a\neq 37$.
\end{itemize}
We recall from Ref.~\cite{Badger:2024fgb} that the MIs \#19 and 20 of ${\rm PB}_B$ belong to the sector which involves a nested square root, whereas the MIs \#35--37 are those of the sector containing elliptic functions.
The 15th MI, on the other hand, is somewhat mysterious: it is in fact canonical on the maximal cut, and its derivatives contain non-logarithmic one-forms not only in correspondence with the elliptic MIs ---~as is expected~--- but also in correspondence with non-elliptic MIs.
It would be interesting to investigate whether this feature can be removed with a basis transformation.

In addition to the DEs, another piece of information we will make use of is the list of which $\eps$-coefficients $\vec{g}_{\rm F}^{\, (k)}(X)$ vanish.
We determine it by performing a number of numerical evaluations of the MIs with \textsc{AMFlow}~\cite{Liu:2017jxz,Liu:2022chg} at random phase-space points.
An important feature is that the MIs of the sectors which are not canonical on the maximal cut (in this case, those in $S \backslash \{ 15\}$) are non-zero only starting from order $\eps^4$.
In other words, they start to contribute only in the finite remainder of the two-loop scattering amplitude.
On the one hand, it is expected that this is the case, as the $\eps$-poles of the amplitudes are determined by the one-loop amplitudes, which are polylogarithmic, as discussed in \cref{sec:struc}.
On the other hand, this is not manifest for an arbitrary choice of MIs, and effort was put in Ref.~\cite{Badger:2024fgb} to satisfy this condition.
Nonetheless, the 15th MI is non-zero already at order $\eps^0$.
We will see below that the non-polylogarithmic one-forms drop out from the solution up to order $\eps^3$ thanks to a conspiracy among the boundary values. 

We are now ready to extend the method of Ref.~\cite{Abreu:2023rco}.
We assume that $\mathrm{F} = {\rm PB}_B$ and neglect the family subscript to simplify the notation.
The same procedure works for ${\rm PB}_B'$ as well.
First, we plug the $\eps$-expansion of \cref{eq:epsexp} into \cref{eq:DEnonCanonical}.
The polynomial $\eps$-dependence of the connection matrix implies an iterative system of DEs for the $\eps$-coefficients,
\begin{align}
\d \vec{g}^{\, (k)}(X) = \Omega^{(0)}(X) \cdot \vec{g}^{\, (k)}(X) + 
   \Omega^{(1)}(X) \cdot \vec{g}^{\, (k-1)}(X) +
      \Omega^{(2)}(X) \cdot \vec{g}^{\, (k-2)}(X)  \,,
\end{align}
for $k \ge 0$, with $\vec{g}^{\, (-2)}(X) = 0 = \vec{g}^{\, (-1)}(X)$.
Next, we drop all $\eps$-coefficients which we have determined numerically to be zero.
We find that the $\vec{g}^{\, (0)}(X)$'s are constant.
Indeed, for all families considered here, $\vec{g}^{\, (0)}$ can be fixed analytically by imposing that $\vec{g}^{\, (1)}(X)$ has the expected branch cuts---i.e., that it contains the expected logarithms. 
This fixes $\vec{g}^{\, (0)}$ up to the normalisation. 
The latter is determined by computing analytically the simplest MI, given by the product of a bubble and a tadpole.
We find that the $\vec{g}^{\, (0)}$'s are rational, in analogy with the polylogarithmic case.\footnote{Strictly speaking, what is necessary for the algorithm to proceed is the knowledge of the linear relations satisfied by the boundary values. Without this information, the solution would contain spurious functions, preventing us from subtracting the poles of the amplitudes analytically.}
Plugging them into the DEs for $\vec{g}^{\, (1)}(X)$ yields a canonical form,
\begin{align} \label{eq:dg1}
\d \vec{g}^{\, (1)}(X) = \sum_i A_{1,i} \, \d \log W_i(X) \cdot \vec{g}^{\, (0)} \,.
\end{align}
We can therefore solve \cref{eq:dg1} in terms of iterated integrals, and substitute the resulting expressions of  $\vec{g}^{\, (0)}$ and $\vec{g}^{\, (1)}(X)$ into the DEs for $\vec{g}^{\, (2)}(X)$.
The derivatives of the 15th MI however appear to contain non-logarithmic one-forms, e.g.
\begin{align}
\d g_{15}^{(2)}(X) = \frac{1}{24} \Bigl[ 12 \, g^{(1)}_{103}(X_0) + 8 \, g^{(1)}_{110}(X_0) + 4 \, g^{(1)}_{111}(X_0) + 
   3 \, g^{(1)}_{118}(X_0) - 48 \, g^{(1)}_{63}(X_0) \Bigl] \, \omega_2(X) + \ldots \,.
\end{align}
This obstruction is removed by a conspiracy of the boundary values: all combinations of boundary values which multiply non-logarithmic one-forms vanish. 
We stress that we do not need to resort to the PSLQ algorithm~\cite{PSLQ} to find such relations among the boundary values.
It suffices to look at the coefficients of the $\omega_i(X)$'s in the derivatives of $g_{15}^{(2)}(X)$, and check numerically that they vanish. 
Interestingly, the constraints on the boundary values obtained this way are a subset of those returned by the algorithm of Ref.~\cite{Abreu:2023rco} applied to $\vec{g}^{\, (1)}(X)$.
Once the non-logarithmic one-forms are removed, the DEs for $\vec{g}^{\, (2)}(X)$ take the canonical form as well, and can be solved in terms of iterated integrals.
We proceed similarly for $\vec{g}^{\, (3)}(X)$ and for the $g^{\, (4)}_i(X)$ with $i \notin S$.
As a result, we have iterated-integral expressions for all $\eps$-coefficients $\vec{g}^{\, (k)}(X)$ with $k=1,2,3$, and at $\eps^4$ for $g^{\, (4)}_i(X)$ with $i \notin S$.

We apply the same procedure to ${\rm PB}_B'$. 
The permutation of the DEs for ${\rm PB}_B$ yields 63 new one-forms, in addition to the 63 defined in Ref.~\cite{Badger:2024fgb}.
Out of these, only 76 are relevant for the solutions truncated at order $\eps^4$.

Finally, we apply the algorithm of Ref.~\cite{Abreu:2023rco} to construct a basis of special functions out of the $\eps$-coefficients for which we have an iterated-integral representation.
These are $\vec{g}_{\rm F}^{(k)}$ for $k \in \{1,2,3,4\}$ and ${\rm F} \in \{ {\rm P}_A, {\rm PB}_A, {\rm PB}_A', {\rm PB}_C \}$, $\vec{g}_{\rm F}^{(k)}$ for $k \in \{1,2,3\}$ and ${\rm F} \in \{ {\rm PB}_B, {\rm PB}_B' \}$,
and $g_{\rm F,i}^{(4)}$ for ${\rm F} \in \{ {\rm PB}_B, {\rm PB}_B' \}$ and $i \notin \{15,19,20,35,36,37\}$.
The resulting number of algebraically independent functions and their breakdown in transcendental weight are shown in \cref{tab:sp_funcs}.

\begin{table}[!h]
\begin{center}
\begin{tabular}{c || c | c | c | c | c || c } 
 transcendental weight & 1 & 2 & 3 & 4 & 4* & all \\
  \hline
\# of functions & 6 & 8 & 45 & 166 & 12 & 237 \\
\end{tabular}
\caption{Number of special functions in our expression of the one- and two-loop amplitudes for $t\bar{t}j$ production at leading colour, broken down by transcendental weight. 
We label by 4* the non-polylogarithmic functions appearing in the MIs at order $\eps^4$.}
\label{tab:sp_funcs}
\end{center}
\end{table}

The 12 remaining $\eps$-coefficients, namely $g_{\rm F,i}^{(4)}$ for ${\rm F} \in \{ {\rm PB}_B, {\rm PB}_B' \}$ and  $i \in \{15,19,20,35,36,37\}$, satisfy coupled DEs involving non-logarithmic one-forms.
Since we cannot solve the latter in terms of iterated integrals, we have no handle over the relations they might satisfy. 
However, it is reasonable to expect that only a limited number of relations can exist among these functions, as they are divided into subsets with different analytic features (the nested square root, the elliptic curve, or both).
Regardless of this, since they are comparatively few and contribute solely to the two-loop finite remainders, we can add them to the generating set of special functions and still achieve the objectives we set at the start of this section.
We dub these functions \emph{non-polylogarithmic}, meaning that they do not have a representation in terms of iterated integrals with logarithmic integration kernels.
We label them by $F^{(4^*)}_i$, for $i=1,\ldots,12$, with an asterisk to remind us that, while these functions appear in the MIs at order $\eps^4$, they do not have a transcendental weight. 

Finally, we have completed the construction of a (potentially over-complete) basis of special functions.
All one- and two-loop MIs relevant for $t\bar{t}j$ production at leading colour are expressed, order by order in $\eps$ up to $\eps^4$, as polynomials in the ring generated by the basis functions and Riemann zeta values ($\zeta_2$ and $\zeta_3$).
For example, we spell out illustrative terms in the expression of the 15th MI of ${\rm PB}_B$:
\begin{align}
\begin{aligned}
g_{15}(X;\eps) = \phantom{ } & \frac{1}{48} + \frac{\eps}{24} \left( F^{(1)}_1 - 2 \, F^{(1)}_2 + 2 \, F^{(1)}_4 - 2 \, F^{(1)}_6 \right) \\
&  - \frac{1}{48} \eps^2 \left[ 7 \, F^{(2)}_1 + 7 \, F^{(2)}_3 + \frac{25}{4} \zeta_2 - \frac{15}{4} \, \left(F^{(1)}_1\right)^2  + 15 \, F^{(1)}_1 \, F^{(1)}_2 - 8 \, \left(F^{(1)}_2 \right)^2 + \ldots \right] \\
& + \frac{1}{8} \eps^3 \left[ F^{(3)}_5 + \frac{820}{9} \zeta_3 + \frac{7}{12} \zeta_2 F^{(1)}_2 - \left(F^{(1)}_2 \right)^3 + 5 \, \left(F^{(1)}_3\right)^2 F^{(1)}_5 + 6 \, F^{(1)}_1 F^{(1)}_2 F^{(1)}_6 + \ldots  \right] \\
& + \eps^4 \, F^{(4^*)}_1 + \mathcal{O}\left(\eps^5\right) \,,
\end{aligned}
\end{align}
where we omit the dependence of $F^{(i)}_j$ on $X$.
As in the fully canonical cases, the special functions inherit from the MIs a parity (either even or odd) with respect to the square roots appearing in the alphabet. 
We list the odd functions and the corresponding square roots in our ancillary files~\cite{ancillary}.

We will see in \cref{sec:res} that, albeit potentially over-complete, the special function basis allows us to subtract the UV/IR poles of the amplitudes analytically, and leads to a dramatic simplification of the expression of the amplitudes with respect to a representation in terms of MIs.
The next subsection will instead be devoted to the numerical evaluation of the special function basis.

We conclude this section by highlighting the conditions underpinning our new method:  
\begin{itemize}
\item the connection matrices are polynomial in $\eps$, and are expressed in terms of $\d \log$'s and $\mathbb{Q}$-linearly independent non-logarithmic one-forms;
\item we have sufficiently many numerical values of the MIs to establish reliably which $\eps$-coefficients vanish and to check the 
relations returned by the algorithm of Ref.~\cite{Abreu:2023rco};
\item the MIs whose DEs are not in canonical form on the maximal cut are non-zero only starting from the highest $\eps$-order that is required for the amplitude computation.
\end{itemize}
The scope of applicability of our new method is therefore significantly larger than that of the previous approach~\cite{Abreu:2023rco}.

\subsection{Numerical evaluation of the special functions}
\label{sec:spfunc_evaluation}

In the final part of this section, we discuss the numerical evaluation of special functions and propose a strategy to address the main bottleneck in this part of the computation: the numerical evaluation of the non-polylogarithmic special functions, $\{F^{(4^*)}_i \}_{i=1}^{12}$, for which no efficient method is currently known.
Our goal is thus to minimise the impact of the non-polylogarithmic functions on the overall numerical evaluation time.
All special functions for which we have a representation in terms of iterated integrals are in fact suitable to be evaluated numerically with the approach discussed in Refs.~\cite{Caron-Huot:2014lda,Gehrmann:2018yef,Chicherin:2020oor}.
The latter is based on constructing a representation in terms of logarithms and dilogarithms for the weight-1 and 2 functions, and one-fold integral representations at weights 3 and 4. 
The one-fold integrations are then performed numerically in a dedicated \textsc{C++} library~\cite{PentagonFunctions:cpp}.
Based on the previous successful applications to two-loop five-particle integrals, it is reasonable to expect that this approach will yield an efficient and stable numerical evaluation.
We leave this to future work, and limit ourselves to giving an explicit expression only for the weight-1 functions:
\begin{align} \label{eq:F2logs}
\begin{aligned}
F^{(1)}_1 & = \log( m_t^2 ) \,,  & F^{(1)}_2 & = \log\bigl[ 2 (d_{12} + m_t^2) \bigr] - \i \pi \,, \quad & F^{(1)}_3 & = \log( 2 d_{23} ) - \i \pi \,, \\
F^{(1)}_4 & = \log( - 2 d_{34} ) \,, \quad  & F^{(1)}_5 & = \log(2 d_{45}) - \i \pi & F^{(1)}_6 & = \log( - 2 d_{15} ) \,,
\end{aligned}
\end{align}
where the logarithms are well-defined in the $45 \to 123$ scattering channel.
We need \cref{eq:F2logs} in order to express the IR/UV subtraction terms in the same special function basis as the amplitudes.
For the non-polylogarithmic functions, we instead construct a minimal system of DEs, which we then solve by means of generalised power series. A detailed performance analysis of this approach is provided at the end of the section.

\begin{figure}
\centering
\begin{tikzpicture}
    \node[anchor=south west, inner sep=0] (pdf) at (0,0) {\includegraphics[scale=0.35]{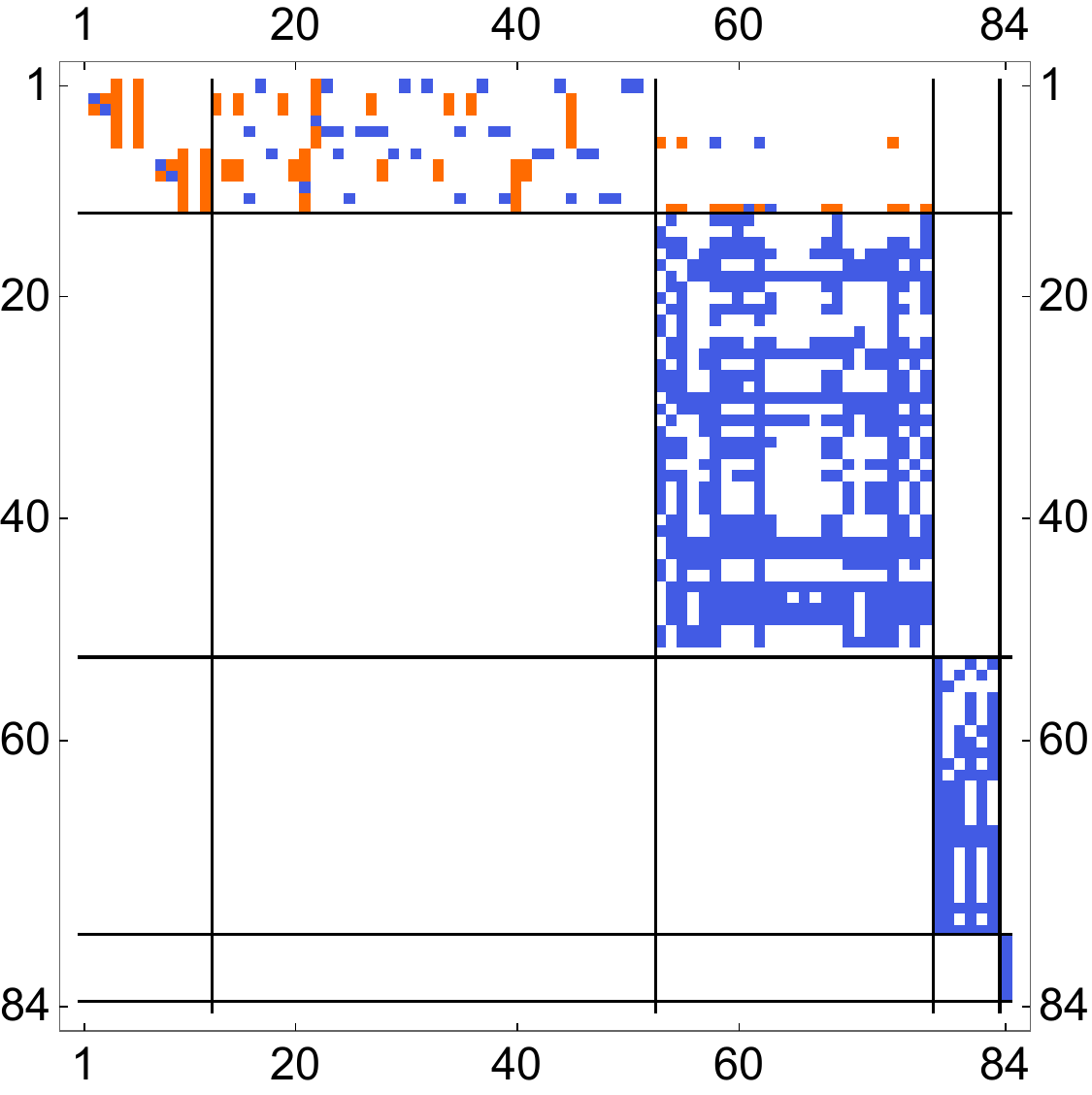}};
    \begin{scope}[x={(pdf.south east)}, y={(pdf.north west)}]
        \draw[decorate, decoration={brace, amplitude=10pt}] (-0.01,0.81) -- (-0.01,0.93)
            node[midway, left=10pt] {$4^*$};
        \draw[decorate, decoration={brace, amplitude=10pt}] (-0.01,0.405) -- (-0.01,0.80)
            node[midway, left=10pt] {$3$};
        \draw[decorate, decoration={brace, amplitude=10pt}] (-0.01,0.155) -- (-0.01,0.395)
            node[midway, left=10pt] {$2$}; 
        \draw[decorate, decoration={brace, amplitude=10pt}] (-0.01,0.09) -- (-0.01,0.145)
            node[midway, left=10pt] {$1$};     
        \draw[->] (-0.12, 0.079) -- (-0.01, 0.079) node[pos=0, left] {$0$};
    \end{scope}
\end{tikzpicture}
\caption{Matrix plot displaying the non-zero entries of the connection matrix in \cref{eq:DEfuncs}.
  Blue dots indicate non-zero entries involving logarithmic one-forms, while the red entries contain also non-logarithmic special functions.
  The solid black lines separate subsets of functions with different transcendental weight, as shown on the left,
  with $4^*$ denoting the non-polylogarithmic special functions.}
\label{fig:connection_matrix}
\end{figure}
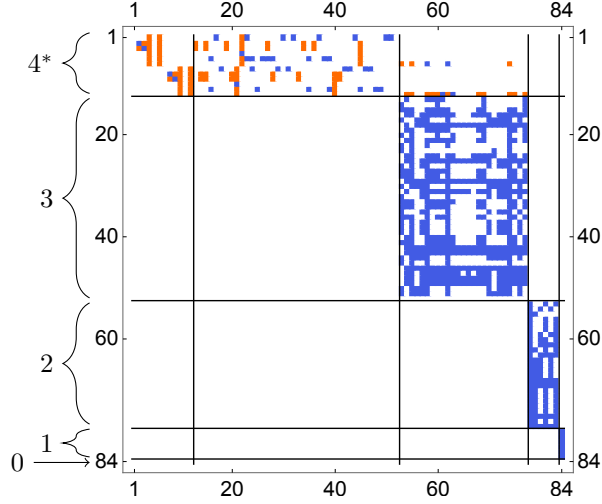

We now focus on the construction of the system of differential equations for the non-polylogarithmic functions $\{F^{(4^*)}_i \}_{i=1}^{12}$~\cite{Badger:2021nhg},
\begin{align} \label{eq:DEfuncs}
 \d \vec{G}(X) = M(X) \cdot \vec{G}(X) \,,
\end{align}
where the connection matrix has the form
\begin{align}
 M(X) = \sum_i A_i \, \d \log W_i(X) + \sum_j B_j \, \omega_j(X) \,.
\end{align}
In \cref{eq:DEfuncs}, $\vec{G}$ is a list of polynomials of special functions and transcendental constants constructed as follows.
We start from the non-polylogarithmic functions.
Next, we add all $\mathbb{Q}$-linearly independent combinations of monomials of special functions required to express the derivatives of the non-polylogarithmic functions, which we know thanks to the DEs for the MIs.
We then take the differential of the newly added functions, and iterate until we reach the differential of weight-1 functions, which requires solely a weight-0 function (chosen to be $1$) to be written.
In addition to the 12 non-polylogarithmic functions, the resulting $\vec{G}$ contains 40, 25, 6 and 1 special function polynomials of weight-3, 2, 1 and 0, respectively, for a total of 84 elements.
For the sake of clarity, we spell out a few terms of $\vec{G}$,
\begin{align}
 \vec{G} = \Bigl(F^{(4^*)}_1 \,,\ldots, F^{(4^*)}_{12} \,, \ldots , F^{(3)}_{39}\,, \ldots ,  F^{(3)}_7 + F^{(1)}_1 F^{(2)}_4 - 2 F^{(1)}_6 F^{(2)}_4 \,, \ldots , \zeta_3 \,, \ldots , F^{(1)}_6 \,, 1 \Bigr)^{\top} \,,
\end{align}
and of the differential of $F^{(4^*)}_1$,
\begin{align} \label{eq:F4*1}
\begin{aligned}
 \d F^{(4^*)}_1 = \phantom{ } & \omega_1 \, F^{(4^*)}_4 + \omega_2 \, F^{(4^*)}_6 - \frac{1}{2} \left(\omega_{31} + \frac{1}{2} \, \d \log W_{69}\right) F^{(3)}_{39}  \\
 & + \frac{1}{4} \, \d \log W_{70} \left( F^{(3)}_7 + F^{(1)}_1 F^{(2)}_4 - 2 \, F^{(1)}_6 F^{(2)}_4 \right) + \ldots \,.
\end{aligned}
\end{align}
We display in \cref{fig:connection_matrix} the non-zero entries of the connection matrix $M(X)$, distinguishing between logarithmic (blue) and non-logarithmic (red) one-forms.
The three addends on the first row on the right-hand side of \cref{eq:F4*1} involve non-logarithmic one-forms and correspond to the red dots on the first row of \cref{fig:connection_matrix}.
The addend on the second row instead corresponds to the first blue dot on the first row of \cref{fig:connection_matrix}.
In \cref{fig:connection_matrix} one can see that the restriction of the connection matrix to the functions with transcendental weight up to 3 is strictly upper triangular, as expected.
The non-polylogarithmic functions are instead coupled to themselves, and in the top left block one can distinguish two sub-blocks associated with the two permutations of ${\rm PB}_B$.
Two functions, $F^{(4^*)}_6$ and $F^{(4^*)}_{12}$, couple also to weight-2 functions; they originate from the 37th MI of ${\rm PB}_B$ and ${\rm PB}_B'$, respectively, whose derivatives are quadratic in~$\eps$.

Finally, we solve the DEs for the special functions $\vec{G}$ by means of generalised power-series expansions~\cite{Moriello:2019yhu}, as implemented in \textsc{DiffExp}~\cite{Hidding:2020ytt}.
This approach has several advantages as compared to solving via generalised power-series the DEs for the MIs.
Firstly, the DEs for the special functions do not depend on $\eps$.
Secondly, while the DEs for the MIs contain information regarding the entire $\eps$ expansion of the solution, the DEs for the special functions contain only the letters and one-forms which are required up to the chosen truncation order.
Finally, the size of the system for the non-polylogarithmic special functions (84) is smaller than the number of MIs of each of the relevant two-loop families (121 each for ${\rm PB}_B$ and ${\rm PB}_B'$), let alone of their union.
These observations lead us to believe that the method we propose represents a substantial improvement over what was previously available in the literature for this class of integrals.

\begin{figure}[t]
    \centering
    \begin{subfigure}[b]{0.32\textwidth}
        \centering
        \includegraphics[height=3.3cm]{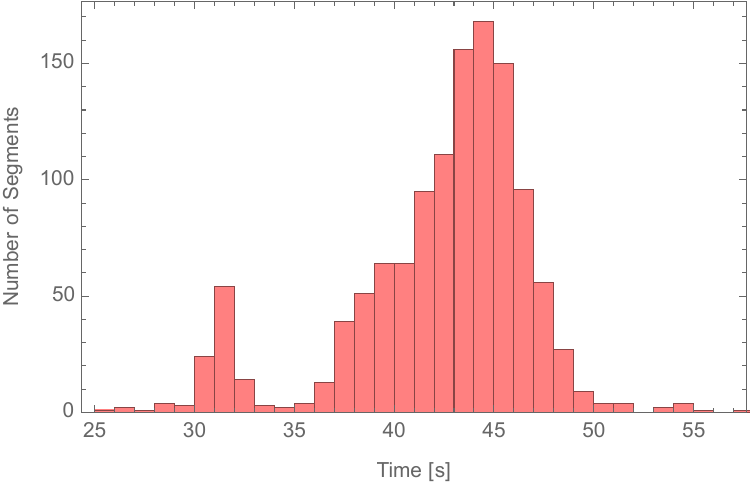}
        \caption{${\rm PB}_A$}
        \label{fig:PBAhisto}
    \end{subfigure}
    \hfill
    \begin{subfigure}[b]{0.32\textwidth}
        \centering
        \includegraphics[height=3.3cm]{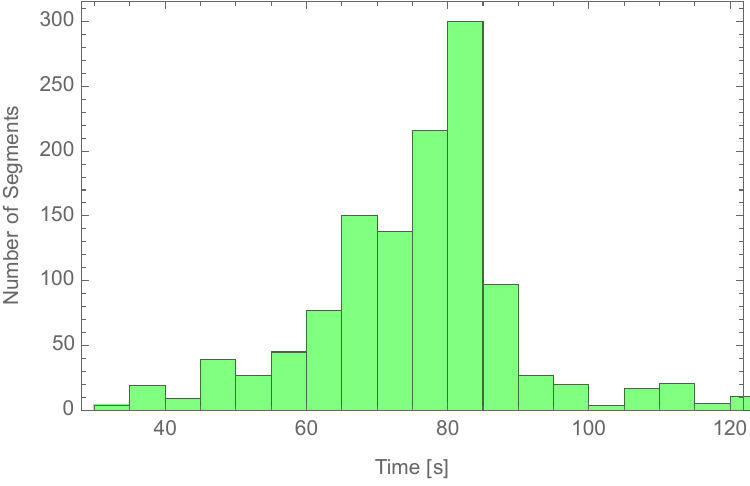}
        \caption{${\rm PB}_B$}
        \label{fig:PBBhisto}
    \end{subfigure}
    \hfill
    \begin{subfigure}[b]{0.32\textwidth}
        \centering
        \includegraphics[height=3.3cm]{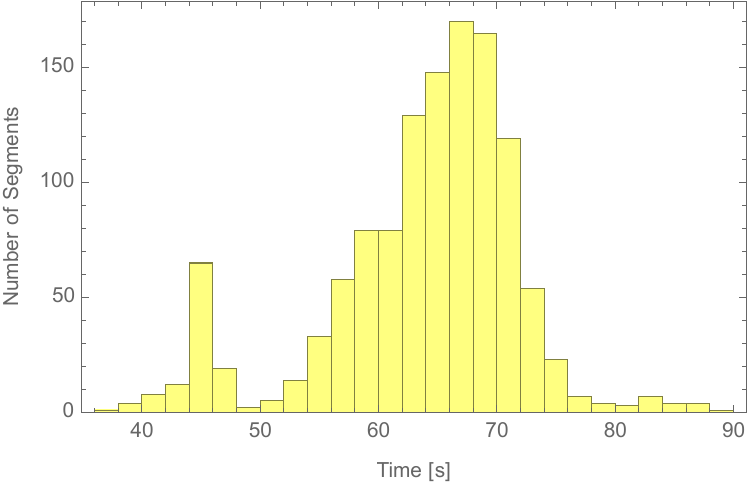}
        \caption{${\rm PB}_C$}
        \label{fig:PBChisto}
    \end{subfigure}

    \caption{Distributions of the evaluation time per segment in the solution of the DEs for the MIs of each family with \textsc{DiffExp}.}
    \label{fig:mishisto}
\end{figure}

\subsection{Performance analysis}
\label{sec:performance}

In the final part of this section, we present benchmark evaluations designed to compare the efficiency of the three different approaches currently available on the market, all making use of generalised power series as implemented in \textsc{DiffExp}: the solution to the DEs for the MIs of each separate family, for the non-polylogarithmic special functions (cf.~\cref{eq:DEfuncs}), and for all special functions.
While the more efficient routine for the polylogarithmic functions is unavailable, we can in fact apply the method discussed above also to all 237 special functions, rather than just the non-polylogarithmic ones.
The dimension of the corresponding system of DEs is 366, with the 12 non-polylogarithmic functions followed by 166, 151, 30, 6 and 1 polylogarithmic functions with transcendental weights 4, 3, 2, 1 and 0, respectively.

\begin{figure}[t]
    \centering
    \begin{subfigure}[b]{0.49\textwidth}
        \centering
        \includegraphics[height=3.3cm]{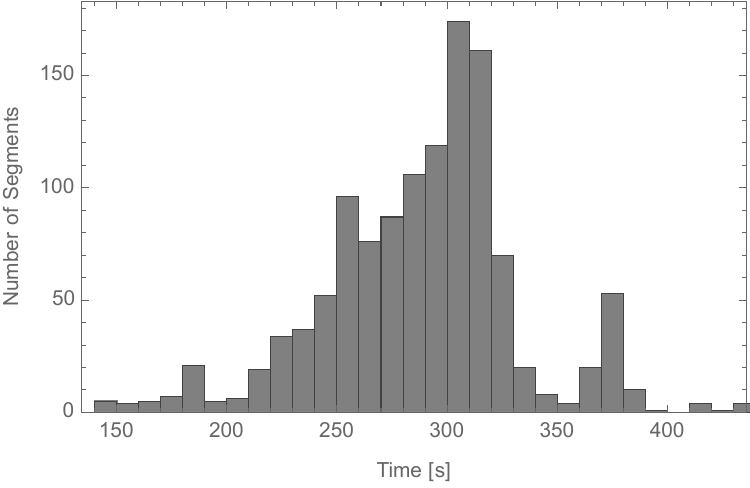}
        \caption{All special functions}
        \label{fig:SFhisto}
    \end{subfigure}
    \hfill
    \begin{subfigure}[b]{0.49\textwidth}
        \centering
        \includegraphics[height=3.3cm]{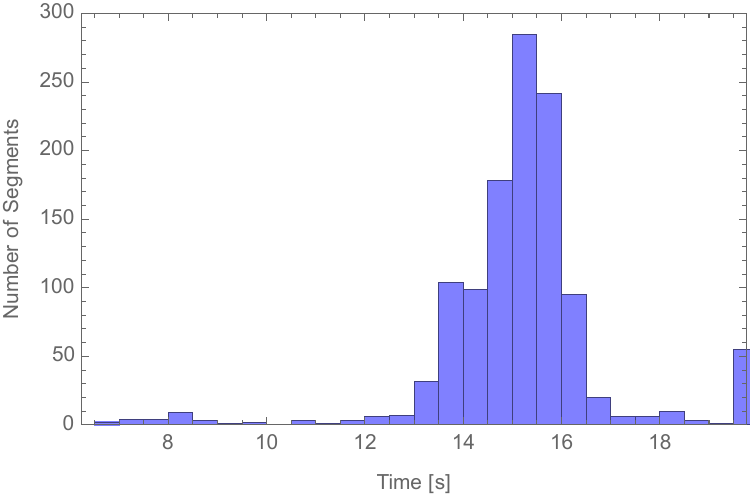}
        \caption{Non-polylogarithmic special functions}
        \label{fig:NonPhisto}
    \end{subfigure}
    \caption{Distributions of the evaluation time per segment in the solution of the DEs for all special functions and for the non-polylogarithmic
    functions alone with \textsc{DiffExp}.}
    \label{fig:Fhisto}
\end{figure}

As a benchmark, we use the average evaluation time per segment in the generalised power series expansion.
We emphasise that the evaluation time per phase-space point strongly depends on the segmentation of the path within the generalised power series expansion method~\cite{Moriello:2019yhu,Hidding:2020ytt}.
An efficient evaluation routine aimed at a large number of phase-space points should thus also aim to minimise the number of segments (see e.g.\ Ref.~\cite{Abreu:2020jxa}).
Therefore, the evaluation time per segment is a more reliable indicator of the method’s performance. The analysis is conducted using a random sample of physical phase-space points in the $s_{45}$ channel,\footnote{All the evaluations are performed on Intel(R) Xeon(R) Gold 6330 CPU @ 2.00GHz with the following \textsc{DiffExp} parameters: \texttt{AccuracyGoal} 16, \texttt{ExpansionOrder} 50, and \texttt{ChopPrecision} 200.} encompassing a total of approximately $1$K segments in each case, starting from the boundary point
\begin{equation} \label{eq:d0}
\vec{d}_0 = \bigl( 2, 1, -1, 5, -2, 1 \bigr)\,.
\end{equation}
We restrict our analysis to points which can be reached from $\vec{d}_0$ by a straight line in the $\vec{d}$-space without exiting the $s_{45}$ channel.\footnote{While it is possible to analytically continue the solution to other regions with \textsc{DiffExp}, this is time consuming and requires more work to determine how the branch cuts have to be crossed. We prefer to restrict ourselves to the $s_{45}$ channel. Points which cannot be reached from $\vec{d}_0$ by a straight line in the $s_{45}$ channel would require piecewise straight paths~\cite{Chicherin:2021dyp} or a dynamic selection of the starting point (see e.g.~\cite{Abreu:2020jxa}). This would not affect our estimate of the timing per segment.}
The values of all MIs at $\vec{d}_0$ are provided in Ref.~\cite{Badger:2024fgb}.
The results on this analysis are summarised in the histograms in \cref{fig:mishisto,fig:Fhisto}.
In \cref{fig:mishisto} we show the distribution of the evaluation time per segment in the solution of the DEs for all sets of MIs, while in \cref{fig:Fhisto} we display the same for the full set of 366 special functions (\cref{fig:SFhisto}), and for the non-polylogarithmic functions alone (\cref{fig:NonPhisto}).
\begin{table}[t]
\centering 
\def\arraystretch{1.1}
\begin{tabular}{|c|c|c|c|c|c|c|}
\hline
 & ${\rm PB}_A$ & ${\rm PB}_B$ & ${\rm PB}_C$ & all MIs & special func.\ (all) & special func.\ (non-polylog.) \\
\hline\hline
$\langle T \rangle$   & $ 43 \, s$ 
					& $77 \, s$ 
					& $66 \, s$
					& $309 \, s$
					& $297 \, s$
					& $16 \, s$ \\
$\sigma$   & $7\,s$
					& $17 \, s$
					& $14 \, s$ 
					& $27 \, s$
					& $65 \, s$
					& $3 \, s$ \\
\hline
\end{tabular}
\caption{Average time per segment and standard deviation for the solution of the DEs for the MIs of each 2-loop family, cumulatively for all 2-loop MIs, for all special functions, and for the non-polylogarithmic special functions alone. The time for all MIs keeps into account that ${\rm PB}_A$ and ${\rm PB}_B$ are needed in two permutations of the external legs, and that the time for the one-loop family (${\rm P}_A$) is $\approx \, 2 s$.}
\label{fig:histodata} 
\end{table}
The average evaluation time per segment and the standard deviation in each case are shown in \cref{fig:histodata}. 
These data do not indicate a significant improvement in the solution of the DEs for all special functions over those for the MIs. 
Indeed, the cumulative evaluation time per segment of the MIs of all relevant families is comparable to that of the special functions. 
However, the average time per segment for the non-polylogarithmic system, of $\approx 16 \, s$, is significantly smaller than that for any set of two-loop MIs. This result confirms the soundness of our strategy. 
Namely, we can reduce substantially the bottleneck of the numerical evaluation of the non-polylogarithmic functions by constructing a minimal system of DEs for them, which are then solved using the generalised power series method. The polylogarithmic functions are instead to be evaluated numerically using the approach discussed in Refs.~\cite{Caron-Huot:2014lda,Gehrmann:2018yef,Chicherin:2020oor,PentagonFunctions:cpp}.

We provide in our ancillary files~\cite{ancillary} 
\begin{itemize}
\item the expression of the master integrals in terms of special functions, 
\item the square-root parity of the special functions, 
\item systems of DEs for both the complete and the non-polylogarithmic sets of special functions, including boundary values at the point in \cref{eq:d0},
\item a \textsc{Mathematica} script to solve the DEs for the special functions using \textsc{DiffExp}.
\end{itemize} 


\section{Results \label{sec:res}}

\begin{center}
\begin{table}
\def\arraystretch{1.3}
\resizebox{\columnwidth}{!}{
\begin{tabular}{|c|c|c|c|c|} 
\hline
Helicity & $R^{(0),[1]}$ & $R^{(0),[2]}$ 
					& $R^{(0),[3]}$ & $R^{(0),[4]}$ \\ 
\hline
$+++$   & $0.26326 -0.0097514 \,\i$ 
					& $0$ 
					& $0$
					& $0$ \\ 
$+-+$   & $5.9619-0.16047 \,\i$
					& $0$
					& $0$ 
					& $-0.31659-0.097935  \,\i$  \\ 
$++-$   & $-5.9575+0.0089231\,\i$ 
					& $-12.606-0.067440 \,\i$ 
					& $ 4.6564 +0.024911 \,\i$ 
					& $-1.9692 -0.010535 \,\i$  \\ 
\hline
\hline
Helicity & $R^{(1),[1]}/R^{(0),[1]}$ & $R^{(1),[2]}/R^{(0),[1]}$ 
					& $R^{(1),[3]}/R^{(0),[1]}$ & $R^{(1),[4]}/R^{(0),[1]}$ \\ 
\hline
$+++$   & $38.396 - 5.8002 \,\i$ 
					& $71.982 - 4.0653 \,\i$ 
					& $-14.289 + 0.70866 \,\i$
					& $17.909 - 0.39528 \,\i$ \\ 
$+-+$   & $19.221 - 8.4151 \,\i$
					& $-4.8506 + 4.8015 \,\i$
					& $0.67096 - 0.09959  \,\i$ 
					& $-1.2201 + 2.1594 \,\i$  \\ 
$++-$   & $20.369 - 19.991 \,\i$ 
					& $41.522 - 41.969 \,\i$ 
					& $-15.990 + 15.739 \,\i$ 
					& $6.2964 - 6.4584 \,\i$  \\ 
\hline
\hline
Helicity & $R^{(2),[1]}/R^{(0),[1]}$ & $R^{(2),[2]}/R^{(0),[1]}$ 
					& $R^{(2),[3]}/R^{(0),[1]}$ & $R^{(2),[4]}/R^{(0),[1]}$ \\ 
\hline
$+++$   & $882.48 - 91.619 \,\i$ 
					& $2489.7 - 266.72 \,\i$ 
					& $-492.28 + 8.1003 \,\i$
					& $593.35 - 87.569 \,\i$ \\ 
$+-+$   & $414.16 - 206.87 \,\i$
					& $-171.78 + 189.69 \,\i$
					& $25.226 - 1.5639 \,\i$ 
					& $-54.820 + 95.716 \,\i$  \\ 
$++-$   & $332.97 - 646.02 \,\i$ 
					& $623.01 - 1325.1 \,\i$ 
					& $-259.14 + 512.33 \,\i$ 
					& $89.185 - 198.65 \,\i$  \\ 
\hline
\end{tabular}
}
\caption{Values of the tree-level helicity sub-amplitudes $R^{(0),[i]}$ and of the finite remainders of the one- and two-loop helicity sub-amplitudes, $R^{(1),[i]}$ and $R^{(2),[i]}$ respectively, as defined by \cref{eq:spindecomposition} with the replacement $A \rightarrow R$. The one- and two-loop sub-amplitudes are normalised by the tree-level sub-amplitude $R^{(0),[1]}$. All sub-amplitudes are evaluated at the phase-space point specified in \cref{eq:evalpoint,eq:tr5choice}. We drop the logarithms of the renormalisation and factorisation scales.}
\label{tab:amplresults}
\end{table}
\end{center}

In this section, we present benchmark values of the two-loop helicity finite remainders for $gg \to t \tb g$ in the leading colour approximation.
These are obtained by putting together the values of the special functions, obtained as discussed in \cref{sec:spfunc_evaluation,sec:performance}, 
and those of the rational coefficients of the finite remainders, resulting from the procedure outlined in \cref{sec:proj}.
In \cref{tab:amplresults} we give the values of the finite remainders of the helicity sub-amplitudes,
\begin{equation}
	R^{(2),[i]}  \qquad \forall \, i \in \lbrace 1, \cdots , 4 \rbrace \,, 
	\nonumber
\end{equation}
obtained by replacing $A$ with $R$ in \cref{eq:spindecomposition} with the same superscripts, and solving for $R^{(2),[i]}$ 
at the selected phase-space point for all the independent helicity configurations of the gluons. We normalised the values by the tree-level sub-amplitude $R^{(0),[1]}$. For completeness, we also provide in \cref{tab:amplresults} the values of the tree-level helicity sub-amplitudes and of the one-loop finite remainders.
In order to facilitate future comparisons, we provide in the ancillary files higher-precision values of the finite remainders, together with the values of the subtracted IR and UV poles needed to recover the mass-renormalised amplitudes~\cite{ancillary}. 

We generated physical phase-space points by parameterising the external momenta in terms of energies and angles, and by sampling randomly the latter. 
Starting from the physical momenta, we compute the corresponding values of the scalar invariants, which we rationalise in order to employ the finite-field setup. The point chosen for the sub-amplitudes evaluation in \cref{tab:amplresults} is given by
\begin{align}
\label{eq:evalpoint}
\begin{alignedat}{3}
	&  d_{12}   = \frac{1617782845110651539}{15068333897971200000} \,, \qquad
	&&d_{23}    = \frac{335}{1232}  \,, \qquad
	&& d_{34}   = -\frac{5}{32}  \,, \\
	& d_{45}    = \frac{3665}{7328} \,, \qquad
	&& d_{15}   = -\frac{45}{1408}  \,, \qquad
	&& m_t^2    = \frac{376940175237098461}{15068333897971200000}  \,, 
\end{alignedat}
\end{align}
with
\begin{equation} \label{eq:tr5choice}
	\mathrm{tr}_5 = \i \, \frac{\sqrt{582950030096630501}}{426229309440} \,.
\end{equation}
The corresponding values of the momentum-twistor variables can be found in our ancillary files~\cite{ancillary}.

The values of special functions and master integrals are cross-checked among the three evaluation strategies discussed in \cref{sec:performance}.
Our results for the rational coefficients are instead validated by verifying the gauge invariance of the amplitudes and by comparing the poles with the predictions from UV renormalisation and IR subtraction as discussed in \cref{sec:struc}. Furthermore, we crosschecked the results presented in \cref{tab:amplresults}---which were obtained using the projector method described in \cref{sec:proj}---against an independent calculation in which we reduce directly the helicity amplitudes in terms of momentum-twistor variables.
For more details on this approach, see Refs.~\cite{Badger:2021owl,Badger:2021imn,Badger:2021ega,Badger:2023mgf,Brancaccio:2024map}.

Although we have not obtained fully analytic results, within the finite-field framework we could already gather some information on the complexity of the rational coefficients of both the master integrals and 
the special function monomials. We observe that the maximum polynomial degree of the rational coefficients goes down by $30\%$
when using our special-function representation of the master integrals with respect to computing the amplitude in terms of master integrals.
This simplification shows the importance of Laurent-expanding the master integrals and expressing the coefficients of the expansion in terms of a basis of special functions.
Furthermore, five weight-4 special functions ($F^{(4)}_i$ with $i \in \{24,34,98,127,151\}$) drop out of the two-loop finite remainders, while all the other special functions---including the non-polylogarithmic ones---appear independently.

\section{Conclusions}

In this article we have computed the two-loop helicity finite remainders for the production of a pair of on-shell top quarks in association with a gluon in gluon fusion ($g g \to t \bar{t} g$) at benchmark physical phase-space points.
The helicity formalism we employed retains the complete information on the spin state of the top quarks, so that their decays can be included straightforwardly in the narrow-width approximation.

In order to achieve this result, we developed a new strategy to express the master integrals in terms of a (potentially over-complete) basis of special functions by solving the associated differential equations without requiring them to be in the canonical form.
Elliptic functions are in fact known to appear in the solution, and a canonical form of the DEs for all two-loop master integrals is not available.
We used numerical evaluations to determine which terms of the master integrals are zero, and exploited this information to extract the polylogarithmic part of the solution to the DEs and construct a basis of special functions to express it using known techniques~\cite{Abreu:2023rco}.
The remaining non-polylogarithmic special functions are treated as independent, but are few and only appear in the finite part of the two-loop amplitudes. 
The latter property is necessary to maintain consistency with the universal pole structure, which cannot include any elliptic functions, but is hidden with an arbitrary choice of master integrals.
Furthermore, we observe that solving the minimal subset of DEs which define the few non-polylogarithmic functions using generalised power series~\cite{Moriello:2019yhu,Hidding:2020ytt} is significantly faster than solving the larger system satisfied by the full set of special functions or by the two-loop master integrals. 
This observation suggests that an efficient numerical evaluation may be achieved as follows. 
The subset of polylogarithmic special functions can be evaluated numerically with the method applied to the pentagon functions~\cite{Caron-Huot:2014lda,Gehrmann:2018yef,Chicherin:2020oor}, which is expected to yield a comparatively negligible evaluation time.
This will require deriving a dedicated representation of the special functions, but the methodology is well understood and we expect the application to be straightforward.
Only for the 12 non-polylogarithmic functions, then, we would need to resort to generalised power series.
We leave to future work the optimisation of this part of the evaluation to target a large number of phase-space points by recycling iteratively the values obtained with previous evaluations (see e.g.\ Ref.~\cite{Abreu:2020jxa}) and tuning the parameters of the algorithm (working precision, expansion order, etc.).

The representation of the master integrals in terms of special functions also leads to a major simplification of the finite remainders, which we evaluated numerically with a routine based on finite-field sampling in \textsc{FiniteFlow} to overcome the high algebraic complexity of the rational coefficients accompanying the special functions. 

This work paves the way to achieve a fully analytical computation of the two-loop amplitudes required to describe $t\bar{t}$ + jet production at the LHC at NNLO in QCD.
Furthermore, we expect that our method to extract the polylogarithmic part of the solution to non-canonical differential equations will find application in other computations in which a canonical form is currently out of reach.

\section*{Acknowledgements}
We thank Christoph Dlapa for many useful discussions.
S.B.\ and C.B.\ acknowledge funding from the Italian Ministry of Universities and Research (MUR) through FARE grant R207777C4R and through grant PRIN 2022BCXSW9.
M.B.~acknowledges funding from the European Union’s Horizon Europe research and innovation programme under the ERC Starting Grant No.~101040760 \emph{FFHiggsTop}.
H.B.H.~has been supported by an appointment to the JRG Program at the APCTP through the Science and Technology Promotion Fund and Lottery Fund of the Korean Government and by the Korean Local Governments~--~Gyeongsangbuk-do Province and Pohang City.
S.Z.~was supported by the European Union’s Horizon Europe research and innovation programme under the Marie
Skłodowska-Curie grant agreement No.~101105486, and by the Swiss National Science Foundation (SNSF) under the Ambizione grant No.~215960.

\appendix

\section{Renormalisation factors and infrared pole operators}
\label{app:renormalisation}

In this appendix we list all terms that are needed to renormalise the amplitude according to \cref{eq:Amren,eq:UVRen} and to cancel the IR singularities as described in \cref{eq:IRRen}. The leading-colour mass counterterms in the integral representation are given by~\cite{Badger:2021owl}
\begin{align} \label{eq:dZm}
\begin{aligned}
	\dZ_m^{(1)} = {} & N_c \; \frac{2+d_s-2d_s \epsilon }{4 (1-2\epsilon) m_t^2} \; I_1 \,, \\
	\dZ_m^{(2)} = {} & N_c^2 \; \bigg\lbrace 
 	  \frac{1}{m_t^4} \left[ \frac{1-6\epsilon+6\epsilon^2}{4(1-2\epsilon)^2}
  	 - d_s \frac{2-3\epsilon}{4(1-2\epsilon)} + \frac{d_s^2}{16} \right] I_2 \\
 	  & + \frac{1}{m_t^2 (1-\epsilon)} \left[ -\frac{7-16\epsilon+12\epsilon^3}{4(1-2\epsilon)(1-4\epsilon)}
 		- d_s \frac{1-8\epsilon+10\epsilon^2}{4(1-4\epsilon)} + d_s^2 \frac{1-2\epsilon}{16} \right] I_3 \bigg\rbrace \,,
\end{aligned}
\end{align}
where $d_s$ is the transverse dimension of the gluon polarisation, and the integrals $I_i$ are defined by
\begin{align} \label{eq:dZmInt}
\begin{aligned}
	I_1 &= \int \dk{1} \; \frac{1}{k_1^2-m_t^2} \,, \\
	I_2 &= \int \dk{1}\dk{2} \; \frac{1}{(k_1^2-m_t^2) (k_2^2-m_t^2)} \,, \\
	I_3 &= \int \dk{1}\dk{2} \; \frac{1}{(k_1^2-m_t^2) k_2^2 (k_1+k_2+p)^2} \,, 
\end{aligned}
\end{align}
with $p^2=m_t^2$.
We set $d_s = 4 - 2\eps$ in the ’t Hooft-Veltman scheme.
The $L$-loop renormalisation constants $\delta{Z}^{(L)}_{t}$~\cite{Melnikov:2000zc} and $\delta{Z}^{(L)}_{\as}$, keeping only the terms that contribute at leading colour and setting $\log(\mu_R)=0$, are given by
\begin{align} \label{eq:ZUV1Loop}
\begin{aligned}
	\delta{Z}^{(1)}_{\as} &= -\frac{\beta_0}{\epsilon} \,, \\
	\delta{Z}^{(1)}_{t} &= C_F \left( -\frac{3}{\epsilon} 
	- \frac{4}{1-2 \epsilon} \right) \,,
\end{aligned}
\end{align}
at one loop, while at two loops we have
\begin{align} \label{eq:ZUV2Loop}
\begin{aligned}
 \dZ^{(2)}_{\as} = {} & \frac{\beta_0^2}{\epsilon^2} - \frac{\beta_1}{2 \epsilon} \,, \\
 \dZ^{(2)}_{t} = {} & C_F^2 \left( \frac{9}{2 \epsilon^2} + \frac{51}{4 \epsilon}
    + \frac{433}{8} - 24 \zeta(3) + 96 \zeta(2) \log(2)-78 \zeta(2) \right) \\
    & + C_F C_A \left( -\frac{11}{2 \epsilon^2} -\frac{101}{4 \epsilon} 
    - \frac{803}{8} +12 \zeta(3) - 48 \zeta(2) \log(2) + 30 \zeta(2) \right) \,.
\end{aligned}
\end{align}
The beta-function coefficients are 
\begin{equation} \label{eq:BetaFunction}
	\beta_0 = \frac{11}{3} C_A, \quad \beta_1 = \frac{34}{3} C_A^2 \,,
\end{equation}
with\footnote{We write $C_F$ in full colour for clarity, although only the leading colour term is needed.}
\begin{equation} \label{eq:Casimir}
	C_A = N_c \,, \qquad C_F = \frac{N_c^2-1}{2 N_c} \,.
\end{equation}
The dependence on $\mu_R$ can be recovered by dimensional analysis.

The pole operator $\ZZ$ appearing in the IR factorisation in \cref{eq:IRRen} is given by~\cite{Becher:2009cu, Becher_2009}
\begin{align} \label{eq:ZIR}
\begin{aligned}
	\ZZ = {} & 1+ \frac{\as}{4\pi} \left( \frac{\GammaPrime_0}{4\eps^2} + \frac{\GammaBold_0}{2\eps} \right) \,, \\
	&+  \left(\frac{\as}{4\pi}\right)^2 \left[ \frac{(\GammaPrime_0)^2}{32\eps^4} 
     + \frac{\GammaPrime_0}{8\eps^3}\bigg(\GammaBold_0 - \frac{3}{2} \beta_0 \bigg) 
     + \frac{\GammaBold_0}{8\eps^2} \big(\GammaBold_0 - 2\beta_0 \big)
     + \frac{\GammaPrime_1}{16\eps^2}
     + \frac{\GammaBold_1}{4\eps} \right] \,. 
\end{aligned}     
\end{align}
The coefficients $\GammaBold_n$ and $\GammaPrime$ are defined through the expansion
\begin{equation} \label{eq:GammaExpansion}
	\GammaBold = \sum_{n \geq 0} \mathbf{\Gamma}_n \bigg(\frac{\as}{4\pi}\bigg)^{n+1} \,, \quad
	\GammaPrime = \sum_{n \geq 0} \mathbf{\GammaPrime}_n \bigg(\frac{\as}{4\pi}\bigg)^{n+1} \,, 
\end{equation}
and can be written in terms of anomalous dimensions,
\begin{equation} \label{eq:AnomDimExpansion}
	\gamma^i(\as) = \sum_{n \geq 0} \gamma^i_n(\as) \bigg(\frac{\as}{4\pi}\bigg)^{n+1} \,, 
\end{equation}
for $i \in \{\mathrm{cusp},g,Q\}$, where $Q$ denotes a massive quark. In this notation, for the $gg \to t \tb g$ amplitude, we have 
\begin{align} \label{eq:GammaN}
\begin{aligned}
	\GammaPrime_n &= -3 \, C_A \gamma^{\mathrm{cusp}}_n \,, \\ 
	\GammaBold_n &= -C_A \frac{\gamma^{\mathrm{cusp}}_n}{2} \bigg(
	\mathcal{L}_{m,23} + \mathcal{L}_{m,15} + \mathcal{L}_{34} + \mathcal{L}_{m,45} \bigg)
	+ 3\gamma^g_n+2 \gamma^Q_n \,. 
\end{aligned}
\end{align}
In the above relations, we have used the following convention for the anomalous dimensions,
\begin{align} \label{eq:AnomDim}
\begin{aligned}
	\gamma^{\mathrm{cusp}}_0 &= 4 \,, \\
	\gamma^{\mathrm{cusp}}_1 &= \left( \frac{268}{9} - \frac{4\pi^2}{3} \right) C_A \,, \\
	\gamma^{g}_0 &= - \beta_0 \,, \\
	\gamma^{g}_1 &= C_A^2 \left(- \frac{692}{27} + \frac{11}{18} \pi^2 + 2 \zeta(3) \right) \,, \\
	\gamma^{Q}_0 &= -2 \, C_F \,, \\
	\gamma^{Q}_1 &= C_F C_A \left( \frac{2\pi^2}{3}-\frac{98}{9}-4 \zeta(3) \right) \,,
\end{aligned}
\end{align}
and the following shorthands for the logarithms,
\begin{align} \label{eq:Logs}
\begin{aligned}
	\mathcal{L}_{ij} &= \log \biggl( \frac{\mu_F^2}{-2d_{ij}} \biggr) \,,  \\
	\mathcal{L}_{m,ij} &= \frac{1}{2} \log \biggl( \frac{\mu_F^2}{-2d_{ij}} \biggr) 
												+ \frac{1}{2} \log \biggl( \frac{m_t^2}{-2d_{ij}} \biggr) \,.
\end{aligned}
\end{align}
As for the renormalisation scale, we also drop all logarithms of $\mu_F$, which can be recovered by dimensional analysis.

\bibliographystyle{JHEP}
\bibliography{ppttj_2L}

\end{document}